\documentclass{JHEP3}
\usepackage{latexsym,amsmath,amssymb}
\pdfoutput=1
\usepackage{graphicx}
\usepackage{eufrak}
\usepackage{slashed}
\usepackage{bm}
\usepackage{epsfig}
\newcommand{\exclude}[1]{}
\usepackage{epsfig}
\usepackage{dcolumn}

\newcommand{\beq}{\begin{equation}}
\newcommand{\eeq}{\end{equation}}
\newcommand{\be}{\begin{eqnarray}}
\newcommand{\ee}{\end{eqnarray}}

\newcommand{\rar}{\rightarrow}

\def\dd{ \,\mathrm{d} }

\def\+{\dagger}
\def\la{\langle}
\def\ra{\rangle}
\def\<{\langle}
\def\>{\rangle}

\newcommand{\Lqcd}{\Lambda_{\mathrm{QCD}}}

\title{The Gauge Fields and  Ghosts in Rindler Space}

\author{  Ariel R. Zhitnitsky\\
Department of Physics \& Astronomy, University of British Columbia, Vancouver, B.C. V6T 1Z1, Canada}

\date{\today}

\abstract{
We consider  2d Maxwell system defined on the Rindler space with metric $ds^2=\exp(2a\xi)\cdot(d\eta^2-d\xi^2)$ with the goal 
to study the dynamics of the ghosts.  We find an extra contribution  to the vacuum energy     in comparison with Minkowski space time
with metric $ds^2= dt^2-dx^2$.
    This extra contribution can be  traced to the unphysical degrees of freedom   (in Minkowski space). 
   The technical reason for this effect to occur  is the property of Bogolubov's coefficients which mix the positive and negative frequencies modes. The corresponding  mixture can not be avoided because  the projections to 
     positive -frequency modes 
  with respect to Minkowski time $t$  and  positive -frequency modes    with respect to the Rindler observer's proper  time $\eta$  are  not equivalent. 
  The exact cancellation of unphysical degrees of freedom which is maintained in Minkowski space can not hold in the Rindler space.
     In BRST approach this effect manifests itself as the presence of BRST charge density    in $L$ and $R$ parts.   An inertial observer in  Minkowski vacuum $ \left| 0 \right>$ observes a universe with no net BRST charge only as a result of cancellation between the two.  However,  the Rindler  observers  who do not ever have access to the entire space time
   would see a net BRST charge. In this respect the effect resembles the Unruh effect.  
The effect is infrared (IR) in nature, and  sensitive to the horizon and/or boundaries.   We interpret the extra energy as the  formation
of the ``ghost condensate" when the ghost degrees of freedom can not propagate, but nevertheless do contribute to the vacuum energy.   Exact computations
in this simple 2d model support the claim made in \cite{UZ} that  the ghost contribution  might be responsible for the observed dark energy    in 4d FLRW universe. }

\begin{document}
\section{Introduction. Motivation}
The main motivation for the present studies is the observation made in \cite{UZ} that the dark energy observed in our universe might be  a result of mismatch between the vacuum energy computed in slowly expanding universe with the expansion rate $H$ (Huble constant) and the one which is computed in flat Minkowski space. If true, the difference between two metrics would lead to an estimate $\Delta E_{vac}\sim H\Lambda_{QCD}^3\sim (10^{-3} {\text eV})^4$ which is amazingly close to the observed value today.

The main idea behind  the claim made in \cite{UZ} can be formulated as follows. It is well known that   in general, in a curved space time it would be not possible to separate positive frequency modes from negative frequency ones in the entire spacetime, in contrast with what happens in Minkowski space where the vector $\partial/\partial t$ is a constant  Killing vector, orthogonal to the $t=\mathrm{const}$ hypersurface.  The Minkowski separation is maintained throughout the whole  space as a consequence of Poincar\'e invariance. It is in a drastic contrast with  a curved space time when 
there are no privileged coordinates.   This means that a transition from a complete orthonormal set of modes to different one (the so-called Bogolubov's transformations) will always mix positive frequency modes  with negative frequency ones.   As a result of this mixture, the vacuum state defined by a particular choice of the annihilation operators will be filled with particles once we switch back to the original basis.   
  Precisely this feature leads to the mismatch $\sim H$  mentioned above between the vacuum energy computed in slowly expanding universe 
  and Minkowski space time.
    
Such drastic, profound consequences arising in going from Minkowski to curved space should not be a surprise to anyone who is familiar with the problem of cosmological particle creation in a gravitational background, or the problem of photon emission by a neutral body which is accelerating. The generic picture   is amazingly simple: the transition from one coordinate system to another leads in general to non-vanishing Bogolubov's coefficients which mix positive and negative frequency modes.  Eventually, it signals a physical production of particles stemming from the interaction with the gravitating background.

The spectrum of the produced particles as well as the rate of production have been discussed in literature in great details\cite{Birrell:1982ix}.  The most important outcome   is  that the typical magnitude of the Bogolubov's coefficients is proportional to the rate at which the background is changing (the Hubble parameter $H$ in case of an expanding universe, or the acceleration rate if we are studying photon emission by a neutral body).
The characteristic frequencies of the gravitationally emitted particles in this set up are of order of the Hubble parameter $\omega_k\simeq H$, whereas higher frequency modes are exponentially suppressed $\sim \exp(-\frac{k}{H})$.    Exactly this feature of the spectrum was a crucial point to identify 
the mismatch energy   $\Delta E_{vac}\sim H\Lambda_{QCD}^3\sim (10^{-3} {\text eV})^4$ with observed dark energy as this type of energy is drastically different from any conventional type of matter. Indeed,  it does not clump because the typical wavelengths $\lambda_k$ of the relevant excitations contributing to  $\Delta E_{vac} $ are of   the order of  entire size of the universe, $\lambda_k\sim k^{-1}\sim H^{-1}\sim 10\text{~Gyr}$.

Precise computations of this sort in a general curved background are difficult to perform. However, as is known, some nontrivial geometrical effects can be explored and understood by analyzing the system that accelerates uniformly with acceleration $a$ through the Minkowski vacuum state, which is the Rindler system.
In this case,  the Bogolubov's coefficients are known to mix the  positive and negative frequency modes. More than that, the  
the Bogolubov's coefficients  exhibit the desired exponential suppression  $\sim \exp(-\frac{k}{a})$ of high frequencies modes. Therefore,   we consider the Rindler space   as a theoretical  laboratory which allows us to understand the dynamics of  gauge theories  in a physically relevant case of expanding universe  when the acceleration parameter $a$ in the flat Rindler space  effectively replaces  the  expansion rate $H$ in non static FLRW universe, while suppressed Bogolubov's coefficients  $\sim \exp(-\frac{k}{a})$ replace $\sim \exp(-\frac{k}{H})$.  

The crucial question we want to address in this work can be formulated as follows. 
It is known that   the fixing a gauge in  the  Lorentz covariant way always lead to emergence of unphysical degrees of freedom which always accompany   the gauge system. The standard way to cure this problem goes back to Gupta and Bleuler formulation  \cite{G, B}  when   the unphysical degrees of freedom (e.g. temporal and longitudinal photons in QED)  drop out of every gauge-invariant matrix element, leaving the theory well defined, i.e., unitary and without negative normed physical states. In particular, the contribution of unphysical degrees of freedom  to the energy momentum tensor vanishes identically in Minkowski space as a result of exact cancellation when appropriate auxiliary   Gupta-Bleuler~\cite{G,B} conditions are imposed. 

We want to see what happens with those unphysical degrees of freedom in accelerating system.  Essentially, we want  to answer the following question: what a Rindler observer   has to say about aforementioned exact cancellation  of  the unphysical degrees of freedom?
We shall see that the accelerating Rindler observer perceives an extra energy   if one compares with the conventional Minkowski vacuum state. More than that, this extra energy can be traced to those unphysical (in Minkowski space) degrees of freedom mentioned above. For conventional massless physical scalar field this effect  is well known as the ``Unruh effect" \cite{Unruh:1976db},\cite{Unruh:1983ms},   and the corresponding physics is  well understood.  
We shall see that the basic reason for the emergence of this extra energy is precisely the same as  for the Unruh effect to occur, and 
it is  resulted from  restriction of Minkowski  vacuum $ \left| 0 \right>$ to the Rindler wedge region 
 where it becomes a thermal state with temperature $T=\frac{a}{2\pi}$. In our case the interpretation is somewhat different as
 we interpret the extra contribution to the energy observed by the Rindler observer as  a result of  formation of a specific   configuration,    the ``ghost condensate" rather than a presence of ``free particles"   prepared in a specific mixed state.
 
 One should emphasize that all these effects  happen to the modes $k\leq a$ when the  entire notion  of ``particle " is not even defined.   In cosmological context when $a\sim H$  and we take    ${k}\sim ~ 1 K\sim 10^{-4}~\text{eV}$ the suppression is of order $\sim \exp(-\frac{{k}}{H})\sim \exp(-10^{27})$ and can be completely ignored for any local related physical phenomena.  The deviations from Minkowski picture start to occur only for modes with very large wave lengths of order size of the universe, $\lambda\sim H^{-1}\sim 10 ~\text{Gyr}$. In different words, the effect is infrared (IR) in nature, and  sensitive to the horizon and/or boundaries.  The phenomenon  does not affect  any local physics.

\section{  Ghosts  dynamics in Minkowski  and in curved spaces. }
In this paper we will be mostly interested in dynamics described by the following lagrangian,
\be\label{lag}
{\cal L} =  {\cal L}_0 + \frac{1}{2} \partial_\mu \phi_2 \partial^\mu \phi_2 - \frac{1}{2} \partial_\mu \phi_1 \partial^\mu \phi_1 , 
\ee
where  $ {\cal L}_0$ describes some physical massive/massless degrees of freedom  which are decoupled and irrelevant for our present study.  This lagrangian emerges in a number of places, such as 2d QED in the chiral limit $m_q=0$ (Schwinger model) as it was formulated by Kogut Susskind \cite{KS}. 
 \footnote{  One should remark that if $m_q\neq 0$ these fields are actually coupled to the physical massive field $\hat\phi  $ as follows  
$m_q \<\bar{q}q\> \cos[ 2\sqrt{\pi} ({\hat\phi + \phi_2 - \phi_1})]$.  However, to simplify things we ignore this interaction in the present discussions.}
The lagrangian (\ref{lag}) also describes photodynamics (when no matter fields are present in the system) where $\phi_1$ and $\phi_2$ are identified with temporal and longitudinal photon's polarizations in any number of dimensions\footnote{Indeed, 
the lagrangian $-\frac{1}{4}F_{\mu\nu}^2$ for the Maxwell field is reduced to the form (\ref{lag}) in the Feynman gauge when the gauge fixing 
term takes the form $-\frac{1}{2}(\partial^{\mu}A_{\mu})^2$   such that the lagrangian describing 
 the non-physical degrees of freedom takes the form 
$\frac{1}{2}(\partial_{\mu}A_1)^2 - \frac{1}{2}(\partial_{\mu}A_0)^2$. In this formula   $A_0$ describes the polarization $\epsilon_{\mu}^{(0)}$ and is identified with the ghost $\phi_1$ in eq. (\ref{lag}), while $A_1$  describes the longitudinal polarization $\epsilon_{\mu}^{(||)}$ in    can be  identified with $\phi_2$.
 The physical, transverse polarizations  $\epsilon_{\mu}^{(\perp)} $ enter the expression for $ {\cal L}_0 $ and decoupled from $\phi_1$ and $\phi_2$.
We should note however, that the decomposition of $A_{\mu}$ field in 2d Schwinger model (when only unphysical polarizations are present in the system)  as adopted  in
\cite{KS} differs from such an identification. }. 
Finally, the same   lagrangian describes the so-called Veneziano ghost\footnote{Not to be confused with conventional Fadeev Popov ghosts.} $\phi_1$ and its partner $\phi_2$ in 4d QCD as discussed in\cite{UZ}. 

One should emphasize  
the  Kogut Susskind (KS) ghost  in 2d (as well as the Veneziano ghost in 4d) we will be dealing with in this paper 
are very different from all other ghosts, including the conventional Fadeev Popov ghosts.
The unique features of KS and the Veneziano  ghosts are due  to their close connections to the  topological properties of the theory. In particular,  the topological density operator $\frac{e}{2\pi} E $ in 2d QED is explicitly expressed in terms of the KS ghost $\phi_1$  as follows, $\frac{e}{2\pi} E= (\frac{e}{2\pi}){\frac{\sqrt{\pi}}{e}}\left(  \Box\hat\phi -  \Box\phi_1 \right)$\cite{KS}.  One should also note that  the appearance of the ghost degree of freedom  in the formalism can be traced  from  conventional Maxwell term $ E^2\sim   \Box^2  $ which contains $\Box^2$ operator.  As is known the $ \Box^2  $ operator can be always re-written in terms of a degree of freedom with a negative kinetic terms.
This is   a simplified explanation how the KS ghost emerges in the system, see \cite{KS} for the details. 
Similar formulae demonstrating the topological features of  the Veneziano ghost also exist in 4d QCD\cite{UZ}. A number of very nontrivial properties    of these ghosts   which will be discussed  in this paper are intimately related to their topological nature. 
\subsection{Conventional picture in Minkowski space}
The most important element for this work  is the presence of the field $\phi_1$ which enters eq. (\ref{lag}) with the negative sign. 
It leads to the following equal time commutation relations needs to be imposed on fields,
\be\label{comm}
\left[ \phi_1 (t, \vec{x})\, , \, \partial_t \phi_1 (t, \vec{y})\right] &=&- i \delta (\vec{x}-\vec{y})\, , \\
\left[ \phi_2 (t, \vec{x})\, , \, \partial_t \phi_2 (t, \vec{y})\right] &=& i \delta (\vec{x}-\vec{y})\, , \nonumber
\ee
The negative sign in eq. (\ref{lag}) however does not lead to any problems when auxiliary (similar to Gupta-Bleuler~\cite{G,B}) conditions on the physical Hilbert space are imposed  by demanding \cite{KS} that the positive frequency part of the free massless combination $(\phi_2 - \phi_1)^{(+)}$ annihilates the physical Hilbert space\footnote{ The original GB subsidiary condition  for 4d QED  are formulated as follows:
$ (\partial^{\mu}A_{\mu})^{(+)} \left|{\cal H}_{\mathrm{phys}}\right> = 0$. In terms of modes this condition takes the form
$(a^{(0)}_k-a^{(||)}_k) \left|{\cal H}_{\mathrm{phys}}\right> = 0$ after the condition $\epsilon_{\mu}^{(\lambda)}k^{\mu}=0$ is imposed. It is 
precisely identical to eq. (\ref{gb1}) after one makes the identification of $A_0$ with $\phi_1$ and $A_1$ with $\phi_2$ as discussed in footnote 2.} :
\be\label{gb}
(\phi_2 - \phi_1)^{(+)} \left|{\cal H}_{\mathrm{phys}}\right> = 0 \, .
\ee
The subsidiary condition~(\ref{gb}) which defines the physical subspace can recast as
\be\label{gb1}
(a_k-b_k) \left|{\cal H}_{\mathrm{phys}}\right> = 0 \, , \;  \< {\cal H}_{\mathrm{phys}}| (a_k^{\dagger}-b_k^{\dagger}) =0 \, ,
\ee
where we expanded $\phi_1$ and $\phi_2$ on a complete orthonormal basis $u_k (t, \vec{x})$ and $v_k(t,\vec{x})$ as
\be
\label{expansion}
\phi_1 (t, \vec{x})&=&\sum_{k}\left[ a_ku_k(t,\vec{x})+a_k^{\dagger}u_k^*(t,\vec{x})\right] \, , \nonumber\\
\phi_2 (t, \vec{x})&=&\sum_{k}\left[ b_kv_k(t,\vec{x})+b_k^{\dagger}v_k^*(t,\vec{x})\right] \, .
\ee
Few comments are in order. Our system is formulated in terms of scalar fields $\phi_1$ and $\phi_2$. But, in fact, this system describes a gauge dynamics, and it is related to the gauge invariance in terms of the original gauge fields as one can see from the construction\cite{KS} for 2d QED, construction~\cite{UZ} for 4d QCD and from footnotes 2, 3 for Maxwell photodynamics in 2d and 4d. Therefore, we treat system (\ref{lag}) as a system which actually describes  the gauge dynamics
when scalar fields $\phi_1$ and $\phi_2$ are treated as auxiliary fields which decouple from physical degrees of freedom  as a result of  subsidiary condition~(\ref{gb}). A related comment is as follows:
  the physical  states which satisfy (\ref{gb}),(\ref{gb1}) are gauge invariant  under {\it positive -frequency gauge transformations} only. This remark will play a crucial role in our following discussions   devoted to analysis of the Rindler states. As we shall see the Rindler states will be invariant under a different set of gauge transformations.  In what follows we shall distinguish the so-called {\it ``proper"} from   {\it ``improper" }gauge transformations when local  gauge invariance is maintained,  while globally it can not hold.  For our specific case presented above:
only positive -frequency gauge transformations that preserve (\ref{gb}) are proper gauge transformations; if the gauge transformations include
a component with a negative frequency mode, it should be treated as ``improper" gauge transformation. 

The equal-time commutation relations~(\ref{comm}) are  equivalent to
\be\label{comm2}
\left[b_k, b_{k'}\right]&=&0 \, , \; [b_k^{\dagger}, b_{k'}^{\dagger}]=0 \, , \; [b_k, b_{k'}^{\dagger}]=\delta_{kk'} \, ,
\ee
for the $\phi_2$ field, whereas for the ghost modes they satisfy
\be
\label{comm1}
\left[a_k, a_{k'}\right]&=&0 \, , \; [a_k^{\dagger}, a_{k'}^{\dagger}]=0 \, , \; [a_k, a_{k'}^{\dagger}]=-\delta_{kk'} \, ,
\ee
where again the sign minus appears in these commutation relations. The ground state $|0\>$ is defined as usual
\be
\label{vacuum}
a_k|0\>=0 \, , ~~~ b_k|0\>=0 \, , ~~~ \forall k \, .
\ee
The sign minus in the commutators~(\ref{comm1}) is known to be carrier of disastrous consequences for the theory if $\phi_1$ is not accompanied by another field $\phi_2$ with properties that mirror and neutralize it.  As thoroughly explained in~\cite{KS}, the condition~(\ref{gb}) or, what is the same,~(\ref{gb1}) are similar to the Gupta-Bleuler~\cite{G,B} condition in QED which ensures that, defined in this way, the theory is self-consistent and unitarity (together with other important properties) is not violated due to the appearance of the ghost.

To see this, one can check that the number operator $\mathrm{N}$ for $\phi_1$ and $\phi_2$ takes the form
\be
\label{N}
\mathrm{N}=\sum_k \left(b_k^{\dagger}b_k- a_k^{\dagger}a_k\right) \, ,
\ee
while the Hamiltonian $\mathrm{H}$ reads
\be
\label{H}
\mathrm{H}=\sum_k \omega_k\left(b_k^{\dagger}b_k- a_k^{\dagger}a_k\right) \, .
\ee
With this form for the Hamiltonian it may seem that the term $- a_k^{\dagger}a_k $ with sign minus implies instability as an arbitrary large number of the corresponding particles can carry an arbitrarily large amount of negative energy.  However,  one can check that  the expectation value for any physical state in fact vanishes as a result of the subsidiary condition~(\ref{gb1}):
\be
\label{H=0}
\< {\cal H}_{\mathrm{phys}}| \mathrm{H} |{\cal H}_{\mathrm{phys}}\>=0 \, .
\ee
In different words, all these ``dangerous'' states which can produce arbitrary negative energy do not belong to the physical subspace defined by eq.~(\ref{gb1}).  The same argument applies to the operator $\mathrm{N}$ with identical result
\be
\label{N=0}
\< {\cal H}_{\mathrm{phys}}| \mathrm{N} |{\cal H}_{\mathrm{phys}}\>=0 \, ,
\ee
where we can see explicitly the pairing and cancelling mechanism at work.
 \subsection{Time dependent background}
  It is well known that there are inherent subtleties and obstacles when one  attempts to formulate a QFT on a curved space with a conventional interpretation of ``particles". As it is known the ``particles"
   is not well defined notion   in a general curved background, see e.g.~\cite{Birrell:1982ix}. 
 In this case there is no a natural choice for the set of modes that on which the fields are expanded, these sets being closely related to a more or less ``natural'' coordinate system.  Indeed, the 
Poincar\'e group is no longer a symmetry of the spacetime and, in general, it would be not possible to separate positive frequency modes from negative frequency ones in the entire spacetime, in contrast with what happens in Minkowski space where the vector $\partial/\partial t$ is a constant a Killing vector, orthogonal to the $t=\mathrm{const}$ hypersurface, and conventional eigenmodes  are eigenfunctions of this Killing vector.  The Minkowski separation is maintained throughout the whole  space as a consequence of Poincar\'e invariance. 

Our goal here is to
compute the contribution of the unphysical modes into the expectation value  (\ref{H=0}) in a curved background. As we mentioned above, the interpretation in terms of particles with specific quantum numbers  (which would be the   canonical way to interpret the results in Minkowski space) can not be given  in this case. However, the computation of the expectation value  (\ref{H=0}) is well posed problem and the answer can be explicitly given in terms of the so-called 
Bogolubov's coefficients, see below. 
 
Therefore, following the standard technique for the computation of particle production in a curved spacetime we consider, along with the expansion~(\ref{expansion}), a second complete set of--barred--modes
\be
\label{expansion2}
\phi_1 (t, \vec{x})&=&\sum_{k}\left[\bar{a}_k\bar{u}_k(t,\vec{x})+\bar{a}_k^{\dagger}\bar{u}_k^*(t,\vec{x})\right] \, ,\\ 
\phi_2 (t, \vec{x})&=&\sum_{k}\left[\bar{b}_k\bar{v}_k(t,\vec{x})+\bar{b}_k^{\dagger}\bar{v}_k^*(t,\vec{x})\right] \, . \nonumber
\ee
The new vacuum state is defined as 
\be
\label{vacuum2}
\bar{a}_k|\bar{0}\>=0 \, , ~~~ \bar{b}_k|\bar{0}\>=0 \, , ~~~ \forall k \, .
\ee
Now, in order to find the contribution of fields  $\phi_1$ and $\phi_2$ into the energy of the ground state, we should expand the new modes
$\bar{u}_k$ and $\bar{v}_k$ in terms of the old ones.  Following the notation of the textbook~\cite{Birrell:1982ix} we obtain
\be
\label{bogolubov}
\bar{u}_k&=&\sum_l \left(\alpha_{kl}u_l+\beta_{kl}u^*_{l}\right) \, ,\\
\bar{v}_k&=&\sum_l \left(\alpha'_{kl}v_l+\beta'_{kl}v^*_{l}\right)\, .  \nonumber
\ee
These matrices are called Bogolubov's coefficients, and they can be evaluated as
\be
\label{bogolubov2}
\alpha_{kl}&=&\left(\bar{u}_k, u_l\right) \, , ~~     \beta_{kl}=-\left(\bar{u}_k, u^*_l\right) \, , \\
\alpha'_{kl}&=&\left(\bar{v}_k, v_l\right) \, , ~~     \beta'_{kl}=-\left(\bar{v}_k, v^*_l\right) \, , \nonumber
 \ee
where the brackets define the generalisation of the conventional scalar product for a curved space
\be
\label{product}
\left(\psi_1, \psi_2\right)=-i\int_{\Sigma}\psi_1(x)\overleftrightarrow{\partial}_{\mu}\psi_2^*\sqrt{-g_{\Sigma}} ~ d\Sigma^{\mu} \, ,
\ee
where $d\Sigma^{\mu}=n^{\mu}d\Sigma$ with $n^{\mu}$ a future-directed unit vector orthogonal to the spacelike hypersurface $\Sigma$ and $d \Sigma$  is the volume element in $\Sigma$.  Any complete set of modes which are orthonormal in the product~(\ref{product}) satisfies
\be
\label{orthonormal}
\left(u_k, u_l\right)=\delta_{kl} \, , ~\left(u^*_k, u^*_l\right)=-\delta_{kl} \, , ~\left(u_k, u^*_l\right)=0 \, ,  \\
\left(v_k, v_l\right)=\delta_{kl} \, , ~\left(v^*_k, v^*_l\right)=-\delta_{kl} \, , ~\left(v_k, v^*_l\right)=0 \, .  \nonumber
\ee
Similar relations, of course, are also valid for the $\bar{u}_k$ and $\bar{v}_k$ modes which appear in the alternative expansion~(\ref{expansion2}).  Equating the two expansions~(\ref{expansion}) and~(\ref{expansion2}) and making use of the orthonormality of the modes~(\ref{orthonormal}), one obtains for the annihilation operators
\be
\label{a}
{a}_k&=&\sum_l \left(\alpha_{lk}\bar{a}_l+\beta_{lk}^*\bar{a}^{\dagger}_{l}\right) \, , \\
{b}_k&=&\sum_l \left(\alpha'_{lk}\bar{b}_l+\beta_{lk}^{'*}\bar{b}^{\dagger}_{l}\right) \, . \nonumber
\ee

The Bogolubov's coefficients possess the set of properties
\be
\label{bogolubov-ortho}
\sum_l \left(\alpha_{lk}\alpha^*_{mk} -\beta_{lk}\beta^*_{mk}\right)&=&\delta_{lm} \, , \\
\sum_l \left(\alpha_{lk}\beta_{mk} -\beta_{lk}\alpha_{mk}\right)&=&0 \, , \nonumber \\
\sum_l \left(\alpha'_{lk}\alpha^{'*}_{mk} -\beta'_{lk}\beta^{'*}_{mk}\right)&=&\delta_{lm} \, , \nonumber \\
\sum_l \left(\alpha'_{lk}\beta'_{mk} -\beta'_{lk}\alpha'_{mk}\right)&=&0 \, . \nonumber
\ee
As one can immediately see from~(\ref{a}), the two Hilbert subspaces based on two possible choices of modes $u_k$ and $v_k$, which appear in~(\ref{expansion}), and $\bar{u}_k$ and $\bar{v}_k$, which instead enter in~(\ref{expansion2}), are different as long as $\beta_{kl}\neq 0, \beta'_{kl}\neq 0$.  In particular, the expectation value of the Hamiltonian~(\ref{H}) of the $k$-th state in the barred vacuum $ \<\bar{0}| \mathrm{H}_k | \bar{0}\> $ is
\be
\label{H2} 
\<\bar{0}|  \omega_k\left(b_k^{\dagger}b_k- a_k^{\dagger}a_k\right)| \bar{0}\>= \omega_k\sum_{l}(|\beta_{kl}|^2+|\beta'_{kl}|^2) \neq 0 \, ,
\ee
which is in sharp contrast with eq.~(\ref{H=0}), derived in Minkowski space.
Few  remarks are in order. \\
$\bullet$ While $a_k^{\dagger}a_k$ partakes in the expression for the Hamiltonian with sign minus, it nevertheless gives a positive sign contribution to the expectation value as a result of an additional minus sign in the commutation relation for the ghost field~(\ref{comm1}).  Hence, no cancellation between the ghost $\phi_1$ and its partner $\phi_2$ could occur in the expectation value~(\ref{H2}), in net contrast with eq.~(\ref{H=0}). The effect is proportional to the Bogolubov's coefficients which mix positive and negative frequency modes. It obviously vanishes when such a mixing does not occur. The effect, however,  does not vanish when it  is   not possible   to separate positive frequency modes from negative frequency ones in the entire space time. \\
\exclude{$\bullet$ One should emphasize that this  effect manifests itself only globally, not locally. Indeed, if we 
define a  subspace of physical states $ |{\cal H}_{\mathrm{phys}}\rangle_{\bar{k}}$ where all modes have momenta $k> \bar{k}\gg H$  such that the notion ``particles"  is  well defined for this subspace, than   the deviation from the standard local physics   will   be  astonishingly small   due to the strong  exponential suppression  $_{\bar{k}} \< {\cal H}_{\mathrm{phys}}| \mathrm{H} |{\cal H}_{\mathrm{phys}}\>_{\bar{k}}
  \sim \exp(-\frac{\bar{k}}{H})$.  Even if we take very small energies  for particles $\bar{k}\sim ~ 1 K\sim 10^{-4}~\text{eV}$ the suppression is of order $\sim \exp(-\frac{\bar{k}}{H})\sim \exp(-10^{27})$ such that  $ _{\bar{k}} \< {\cal H}_{\mathrm{phys}}| \mathrm{H} |{\cal H}_{\mathrm{phys}}\>_{\bar{k}}  =0+{\cal{O}} ( \exp(-10^{27}))$ which is indistinguishable  from the Minkowski space time  result (\ref{H=0}). The   modes which are strongly affected are the ones with 
 wave lengths $\lambda$ of  order of the size of the entire universe $\lambda\sim k^{-1}\sim H^{-1} \sim 10~ \text{Gyr}$. Therefore, the effect manifests itself only on global, not local,  scales.\\}
$\bullet$ The deviation of the expectation value from zero (\ref{H2}) due to the unphysical (in Minkowski space) modes should not be interpreted in terms of particles as entire notion of ``particle" is not well defined for $k\leq H$ where the effect is  pronounced. 
  This is a common  problem of interpretation in terms of particles in a curved background,  and we shall not comment on this problem 
  referring to the textbook \cite{Birrell:1982ix}. We interpret the result (\ref{H2}) as an emergence of an additional contribution to the vacuum energy  in time-dependent background in comparison with  Minkowski space-time.  Any  details about  particles's  quantum numbers   can not be specified as this would require a detector with a size of entire universe  $L\sim k^{-1}\sim H^{-1} \sim 10~ \text{Gyr}$.  Due to the same reason a number of  other related questions (such as negative norm states, unitarity etc) can not be even properly posed as  notion of ``particle" is not well defined for such long wave lengths.  \\
$\bullet$  If we had started with a conventional scalar 
field $\phi_2$ with a positive sign for the kinetic term in eq. (\ref{lag}), without mentioning that
the field from eq.  (\ref{lag}) is actually related to the gauge dynamics   describing  an unphysical degree of freedom (in Minkowski space),
 we would unambiguously predict there existence of  extra energy  given by eq. (\ref{H2}).
  Such an interpretation would be absolutely conventional and commonly  accepted by the  community\cite{Birrell:1982ix}. Some doubts only occur  when one recalls that the field $\phi_2$ was actually unphysical degree of freedom in Minkowski space (it did not belong to the physical Hilbert space as discussed in the text, see eqs.(\ref{gb}), (\ref{gb1})), and therefore, a deeper understanding what is really happening is needed in this case.
  
  To clarify all these (and related) questions we consider  exactly solvable model (2d Maxwell system) using  two drastically different metrics to discuss the dynamics of the gauge fields: 1)   conventional Minkowski  metric $ds^2= dt^2-dx^2$ ~and 2)  the Rindler metric   $ds^2=\exp(2a\xi)\cdot(d\eta^2-d\xi^2)$.  To understand the gauge dynamics in these circumstances and to get a complementary  picture we quantize our system using two approaches. First we use conventional GB condition (\ref{gb}) to select the physical Hilbert space. Secondly, we use BRST operator approach such that we can interpret the emergence of the extra energy (\ref{H2}) from a different perspective  in terms of behaviour of BRST operator.  
  \section{Gauge dynamics in Rindler space}
  Our goal here is to understand the extra energy discussed above (\ref{H2}) by considering the Rindler observer. We shall explicitly compute 
  the Bogolubov's coefficients  in eq. (\ref{H2}) and demonstrate   that the effect is present even for this flat (but still nontrivial) metric.
  One can explicitly see why the cancellation between $\phi_1$ and $\phi_2$ fields which was in effect  in Minkowski space, does not hold for the Rindler observer any more. The crucial difference between the two cases  is:
   the physical  states which are selected by eq. (\ref{gb}) are   gauge invariant  states under  positive -frequency gauge transformations while the Rindler states are the gauge  invariant states under a different set of gauge transformations.  Furthermore, the Rindler  observers  do not ever have access to the entire spacetime because they accelerate   to never enter the forward (backward) lightcones of some events. This is precisely the reason why
   the cancellation (\ref{H=0}) which is maintained in  Minkowski space can  not hold for the Rindler observer.

   We follow notations   \cite{Birrell:1982ix} in our analysis
   and separate  the space time into four quadrants $F, P, L$ and $R$.  We will choose the origin such that these regions are defined by $t>|x|$, $t<-|x|$, $x<-|t|$ and $x>|t|$ respectively.  While no single region contains a Cauchy surface, the union of the left and right regions $L$ and $R$ plus the origin does contain many Cauchy surfaces, for example $t=0$.   
We will write the Minkowski metric with the sign convention
\beq
ds^2=dt^2-dx^2,
\eeq
and the wave equation which follow from (\ref{lag}) possesses standard orthonormal mode solutions  
\beq
\label{M}
u_k=\frac{1}{\sqrt{4\pi\omega}} e^{-i\omega t+ikx}.
\eeq
In the quadrant $R$, called  the right Rindler wedge, one may define the coordinates $(\xi^R,\eta^R)$ via the transformations
\beq
\label{eta}
t=\frac{e^{a\xi^R}}{a}{\text{sinh}}\ a\eta^R, ~~~~~  x= \frac{e^{a\xi^R}}{a}{\text{cosh}}\ a\eta^R
\eeq
where $a$ is a dimensional  constant. We may define coordinates $(\xi^L,\eta^L)$ in the left Rindler wedge $L$ in a similar way with the signs of both $t$ and $x$ reversed~\cite{Birrell:1982ix}.   
In these new coordinates the metric is conformal to the Minkowski metric
\beq
ds^2=e^{2a\xi}(d\eta^2-d\xi^2)
\eeq
and so the positive frequency plane waves will  be of the form 
\be
\label{R}
^Ru_k= \frac{1}{\sqrt{4\pi\omega}} e^{ik\xi^R-i\omega\eta^R}~~~~ {\rm in~ R} ,~~~ ^Ru_k=0~~~~~ {\rm in~ L}  
\ee
\be
\label{L}
^Lu_k=\frac{1}{\sqrt{4\pi\omega}} e^{ik\xi^L+i\omega\eta^L} ~~~~~~{\rm in~ L} , ~~~^Lu_k=0 ~~~~~~~{\rm in~ R} 
\ee
The set ({\ref{R}) is complete in region R, while (\ref{L}) is complete in L, but neither is complete in on all of Minkowski space. 
However, both sets together are complete. 
The sign difference corresponds to the fact that  a right moving wave in R moves towards increasing value of $\xi$, while in L it moves toward decreasing value of $\xi$.  In any case, these modes are positive frequency modes with respect to the time-like Killing vector $+\partial_{\eta}$ in R and  $-\partial_{\eta}$ in L.   
 No linear combination of these two plane waves is holomorphic at the origin, however the sum of the plane wave on one side and $e^{-\pi\omega/a}$ times the conjugate plane wave with negative  wavenumber on the other side is everywhere holomorphic\cite{Unruh:1976db}.

Therefore, for the second complete set of barred modes (\ref{expansion2}) one can use modes (\ref{R}) and (\ref{L}) as follows
\be
\label{expansion-R}
\phi_1=\sum_k\frac{1}{\sqrt{4\pi\omega}}(a^L_ke^{ik\xi^L+i\omega \eta^L}+a^{L\dagger}_ke^{-ik\xi^L-i\omega \eta^L}+a^R_ke^{ik\xi^R-i\omega \eta^R}+a^{R\dagger}_ke^{-ik\xi^R+i\omega \eta^R}) \\ \nonumber
\phi_2=\sum_k\frac{1}{\sqrt{4\pi\omega}}(b^L_ke^{ik\xi^L+i\omega \eta^L}+b^{L\dagger}_ke^{-ik\xi^L-i\omega \eta^L}+b^R_ke^{ik\xi^R-i\omega \eta^R}+b^{R\dagger}_ke^{-ik\xi^R+i\omega \eta^R})
\ee
The Rindler vacuum state is defined as
\be
\label{vacuum-R}
a_k^L|0_R\>=0 \, , ~~~a_k^R|0_R\>=0 \, , ~~~ b_k^L|0_R\>=0 \, , ~~~  b_k^R|0_R\>=0 \, , ~~~ \forall k \, .
\ee
The simplest way to compute the corresponding Bogolubov's coefficients is to note that although $^Ru_k $ and $^Lu_k$ are not analytic, the two combinations 
\be
\label{analytic}
 \exp{(\frac{\pi\omega}{2a})} ~ ^Ru_k  + \exp{(-\frac{\pi\omega}{2a})}~   ^Lu_{-k}^*  \\ \nonumber
  \exp{(-\frac{\pi\omega}{2a})}~   ^Ru_{-k}^*   + \exp{(\frac{\pi\omega}{2a})}~ ^Lu_k 
 \ee
 are analytic and bounded\cite{Unruh:1976db}.  These modes share the positivity frequency analyticity properties of the Minkowski modes (\ref{M}), than they must also share a common vacuum state, see below precise definition. Therefore, instead of expansion (\ref{expansion}) with modes (\ref{M}) we can expand $\phi_1$  in terms of (\ref{analytic}) as
\be 
\label{expansion-M1}
\phi_1&=&\sum_k\frac{1}{\sqrt{ 4\pi\omega}}  \cdot \frac{1}{\sqrt{(e^{\pi\omega/a}-e^{-\pi\omega/a})}} \Big[a^1_k(e^{\frac{\pi\omega}{2a}+ik\xi^R-i\omega\eta^R}+e^{\frac{-\pi\omega}{2a}+ik\xi^L-i\omega\eta^L})  \nonumber \\
&+&a^2_k(e^{\frac{\pi\omega}{2a}+ik\xi^L+i\omega\eta^L}+e^{\frac{-\pi\omega}{2a}+ik\xi^R+i\omega\eta^R})  \nonumber \\
&+&a^{1\dagger}_k(e^{\frac{\pi\omega}{2a}-ik\xi^R+i\omega\eta^R}+e^{\frac{-\pi\omega}{2a}-ik\xi^L+i\omega\eta^L})\nonumber\\
&+&a^{2\dagger}_k(e^{\frac{\pi\omega}{2a}-ik\xi^L-i\omega\eta^L}+e^{\frac{-\pi\omega}{2a}-ik\xi^R-i\omega\eta^R})\Big].
\ee
the same can be done for  $\phi_2$ field:
\be 
\label{expansion-M2}
\phi_2&=&\sum_k\frac{1}{\sqrt{ 4\pi\omega}}  \cdot \frac{1}{\sqrt{(e^{\pi\omega/a}-e^{-\pi\omega/a})}} \Big[b^1_k(e^{\frac{\pi\omega}{2a}+ik\xi^R-i\omega\eta^R}+e^{\frac{-\pi\omega}{2a}+ik\xi^L-i\omega\eta^L})  \nonumber \\
&+&b^2_k(e^{\frac{\pi\omega}{2a}+ik\xi^L+i\omega\eta^L}+e^{\frac{-\pi\omega}{2a}+ik\xi^R+i\omega\eta^R})  \nonumber \\
&+&b^{1\dagger}_k(e^{\frac{\pi\omega}{2a}-ik\xi^R+i\omega\eta^R}+e^{\frac{-\pi\omega}{2a}-ik\xi^L+i\omega\eta^L})\nonumber\\
&+&b^{2\dagger}_k(e^{\frac{\pi\omega}{2a}-ik\xi^L-i\omega\eta^L}+e^{\frac{-\pi\omega}{2a}-ik\xi^R-i\omega\eta^R})\Big], 
\ee
where $ b_k^1, b_k^2$ satisfy the following commutation relations,
\be\label{comm_2}
\left[b_k^{(1,2)}, b_{k'}^{(1,2)}\right]&=&0 \, , \; [b_k^{{(1,2)}\dagger}, b_{k'}^{{(1,2)}\dagger}]=0 \, , \; [b_k^{(1,2)}, b_{k'}^{{(1,2)}\dagger}]=\delta_{kk'} \, ,
\ee
 whereas $a_k^1, a_k^2$ for the ghost field $\phi_1$  satisfy
\be
\label{comm_1}
\left[a_k^{(1,2)}, a_{k'}^{(1,2)}\right]&=&0 \, , \; [a_k^{{(1,2)}\dagger}, a_{k'}^{{(1,2)}\dagger}]=0 \, , \; [a_k^{(1,2)}, a_{k'}^{{(1,2)}\dagger}]=-\delta_{kk'} 
\ee
where again the sign minus appears in these commutation relations. The
 Minkowski vacuum state is determined as usual 
\be
\label{vacuum-M}
a_k^1|0\>=0 \, , ~~~a_k^2|0\>=0 \, , ~~~ b_k^1|0\>=0 \, , ~~~  b_k^2|0\>=0 \, , ~~~ \forall k \, .
\ee
This equation   replaces eq. (\ref{vacuum}).
  Matching coefficients in (\ref{expansion-R})  with (\ref{expansion-M1}) and  (\ref{expansion-M2})  one finds the Bogoliubov's coefficients\cite{Birrell:1982ix,Unruh:1976db}, 
\be
\label{Bogolubov}
a^L_k=\frac{e^{-\pi\omega/2a}a^{1\dagger}_{-k}+e^{\pi\omega/2a}a^2_k}{\sqrt{e^{\pi\omega/a}-e^{-\pi\omega/a}}}~~~~~~
a^R_k=\frac{e^{-\pi\omega/2a}a^{2\dagger}_{-k}+e^{\pi\omega/2a}a^1_k}{\sqrt{e^{\pi\omega/a}-e^{-\pi\omega/a}}}\\  
b^L_k=\frac{e^{-\pi\omega/2a}b^{1\dagger}_{-k}+e^{\pi\omega/2a}b^2_k}{\sqrt{e^{\pi\omega/a}-e^{-\pi\omega/a}}}~~~~~~
b^R_k=\frac{e^{-\pi\omega/2a}b^{2\dagger}_{-k}+e^{\pi\omega/2a}b^1_k}{\sqrt{e^{\pi\omega/a}-e^{-\pi\omega/a}}}. \nonumber
\ee
Now consider an accelerating Rindler observer at $\xi= $const. As is known, such an observer's proper time is proportional to $\eta$. The vacuum for this observer is determined by (\ref{vacuum-R}) as this is the state associated with the positive frequency modes with respect to $\eta$.
A Rindler observer in (R,L)  will measure the energy using  the Hamiltonian $\mathrm{H}^{(R,L)}$ which is given by
\be
\label{H-R}
\mathrm{H}^{(R,L)}=\sum_k \omega_k\left(b_k^{(R,L)\dagger}b_k^{(R,L)}- a_k^{(R,L)\dagger}a_k^{(R,L)}\right) \, .
\ee
The subsidiary condition~(\ref{gb}) defines the physical subspace for accelerating Rindler observer 
\be\label{gb-R}
\left(a_k^{(R,L)}-b_k^{(R,L)}\right) \left|{\cal H}_{\mathrm{phys}}^{(R,L)}\right> = 0 \, ,  
\ee
such that the exact cancellation between $\phi_1$ and $\phi_2$ fields holds  for any physical state defined by eq. (\ref{gb-R}), i.e.
\beq
  \left<{\cal H}_{\mathrm{phys}}^{(R,L)}|\mathrm{H}^{(R,L)}  |{\cal H}_{\mathrm{phys}}^{(R,L)}\right> = 0 
  \eeq
as it should.
However, if the system is in the Minkowski vacuum state $ |0\> $ defined by (\ref{vacuum-M}) a Rindler observer using the same Hamiltonian 
(\ref{H-R}) will observe the following amount  of energy in mode $k$, 
\be
\label{RR}
\< 0 | \omega_k\left(b_k^{(R,L)\dagger}b_k^{(R,L)}- a_k^{(R,L)\dagger}a_k^{(R,L)}\right)  |0\>= 
  \frac{2\omega e^{-\pi\omega/a}}{{(e^{\pi\omega/a}-e^{-\pi\omega/a})}}= \frac{2\omega }{(e^{2\pi\omega/a}-1)}.
\ee
This  is the central result of this section and is a direct analog of  eq. (\ref{H2})  discussed previously.
In the present, exactly solvable model, one can explicitly see the nature of this non-cancellation between two unphysical fields 
as the Bogolubov's coefficients can be exactly computed in this case. In fact, one can construct 
 the Minkowski vacuum state $ |0\> $ in terms of the Rindler's states, the so-called `` squeezed state", see Appendix A for the details.
Few remarks are in order:\\
$\bullet$  As we mentioned earlier, if we had started with a conventional scalar 
field $\phi_2$ with a positive sign for the kinetic term  
the result (\ref{RR}) would represent a well-known effect on the Plank spectrum for radiation at $T=a/(2\pi)$, see \cite{Birrell:1982ix,Unruh:1976db}
with the only difference that we have extra degeneracy factor 2 as a result of two fields $\phi_2$ and $\phi_1$ instead of one field. Our fields, however, are related to unphysical (in Minkowski space) degrees of freedom. Therefore, the result (\ref{RR}) is quite unexpected. \\
$\bullet$  No cancellation between the ghost $\phi_1$ and its partner $\phi_2$ could occur in the expectation value~(\ref{RR}), in net contrast with eq.~(\ref{H=0}) as  a result of opposite sign in commutator (\ref{comm_1}) along with negative sign in Hamiltonian (\ref{H-R}).  \\
  $\bullet$   The contribution  of higher frequency modes are exponentially suppressed $\sim \exp(-\omega/a)$ as expected. 
 The interpretation of eq. (\ref{RR}) in terms of particles is very problematic (as usual for such kind of problems)  as typical frequencies  when the effect (\ref{RR}) is not exponentially small, are of order $\omega\sim a$, and notion of ``particle" for such $\omega$  is not well defined. 
  In addition, in order to properly interpret this extra contribution  (\ref{RR}) one should consider the  particle detector moving along the world line,
    see section 
  \ref{interpretation} and Appendix B for details.
 \\
 $\bullet$ Let us   define a  subspace of physical states $ |{\cal H}_{\mathrm{phys}}\rangle_{\bar{k}}$ where all modes have momenta $k> \bar{k}\gg a$  such that the notion ``particles"  becomes   well defined for this subspace. For these states  
 the deviation from the standard local physics   will   be  astonishingly small   due to the strong  exponential suppression  $_{\bar{k}} \< {\cal H}_{\mathrm{phys}}| \mathrm{H} |{\cal H}_{\mathrm{phys}}\>_{\bar{k}}
  \sim \exp(-\frac{2\pi\bar{k}}{a})$.  In cosmological context when $a=H$  and we take    $\bar{k}\sim ~ 1 K\sim 10^{-4}~\text{eV}$ the suppression is of order $\sim \exp(-\frac{\bar{k}}{H})\sim \exp(-10^{27})$ such that  $ _{\bar{k}} \< {\cal H}_{\mathrm{phys}}| \mathrm{H} |{\cal H}_{\mathrm{phys}}\>_{\bar{k}}  =0+{\cal{O}} ( \exp(-10^{27}))$ which is indistinguishable  from the Minkowski space time  result (\ref{H=0}).     The deviations from Minkowski picture start to occur only for modes with very large wave lengths $\lambda\sim a^{-1}$ for  small $a$. In different words, the effect is infrared (IR) in nature, and  sensitive to the horizon and/or boundaries. The conventional local physics with   $k> \bar{k}\gg a$ is not affected by unphysical (in Minkowski space) degrees of freedom with very high degree of accuracy. 
  Such a sensitivity to IR physics is obviously related to the topological nature of the KS ghost, and will be discussed in details in section \ref{results}.
   \\
 $\bullet$ One can explicitly see why the cancellation (\ref{H=0}) of unphysical degrees of freedom in Minkowski space fail  to hold 
 for the accelerating  Rindler observer (\ref{RR}).  
  The   selection of  the physical Hilbert subspace  (\ref{gb1}) is based on the properties of the operator which selects   positive -frequency modes 
  with respect to Minkowski time $t$. At the same time the Rindler observer selects the physical Hilbert space (\ref{gb-R}) by using positive -frequency modes    with respect to observer's proper  time $\eta$. These two sets are obviously not equivalent, as e.g. they represent a mixture of 
  positive and negative frequencies modes defined in R- and L- Rindler wedges.  At the same time, the Rindler  observers  do not ever have access to the entire space time. Therefore, from the Rindler's view point the cancellation in Minkowski space can be only achieved if one uses both sets (L and R). Of course, using the both sets would contradict to  the basic principles as the R-Rindler observer does not have access to the L wedge even  for arbitrary small  acceleration parameter $a$. \\
  $\bullet$ One should also recall that our system is actually originated from a gauge invariant  QFT. More than that,  the selection of gauge  invariant sector of the theory is  formulated in terms of  positive -frequency  operator $ (\partial^{\mu}A_{\mu})^{(+)} \left|{\cal H}_{\mathrm{phys}}\right> = 0$ which reduces to (\ref{gb}), see footnote 4 on pg.4. The selection of gauge invariant sectors is obviously different whether one uses  Minkowski time $t$ or the Rindler observer's proper  time $\eta$ for selecting the positive frequency operator. As we mentioned above  this difference does not affect any local physics when one deals with physical subspace $ |{\cal H}_{\mathrm{phys}}\rangle_{\bar{k}}$, but it does change the IR physics at very large distances $\sim a^{-1}$   which plays the role of the inverse Hubble constant $H^{-1}\sim 10~ \text{Gyr}$ for FLRW universe. 
  
  Finally, to simplify things, we formulated our problem in terms of the scalar unphysical degrees of freedom $\phi_1$ and $\phi_2$   in 2 dimensions as well as in 4d QCD~\cite{UZ}. However, the same problem can be treated directly in 4 dimensions by using  conventional $4-$ vectors 
  instead of their scalar  components expressed by $\phi_1$ and $\phi_2$ fields. The corresponding computations\cite{ohta} confirm our findings. 
  
  To elaborate on these important points we shall study in next section the same system using BRST quantization  for  selection of  the physical Hilbert 
  space (instead of Gubta- Bleuler formulation 
  exploited  in this section). 
  We shall see how the effect discussed in this section manifests itself  in term of    global properties of BRST operator.
    \section{BRST in Rindler space.}
    The BRST quantization\footnote{For a general introduction into the BRST technique, see e.g. \cite{Weinberg}.} in the Rindler space has been discussed previously in the literature~\cite{Evslin:2005zv}. While we agree with technical details  of ref. \cite{Evslin:2005zv}, our interpretation of the obtained results is quite different. 
    
 We start with   
the   lagrangian $-\frac{1}{4}F_{\mu\nu}^2$ for the Maxwell field. In the Feynman gauge we add  the gauge fixing 
term $-\frac{1}{2}(\partial^{\mu}A_{\mu})^2$   such that the lagrangian describing 
 the non-physical degrees of freedom takes the form 
$\frac{1}{2}(\partial_{\mu}A_1)^2 - \frac{1}{2}(\partial_{\mu}A_0)^2$. In this formula   $A_0$ describes the polarization $\epsilon_{\mu}^{(0)}$ and is identified with the ghost $\phi_1$ in eq. (\ref{lag}), while $A_1$  describes the longitudinal polarization $\epsilon_{\mu}^{(||)}$ in 2d QED and  is identified with $\phi_2$. In BRST approach we must also add the $c-$ ghost field which is anti commuting scalar field such that final lagrangian 
 to be  studied in this section takes the form
\be
\label{lag_BRST}
 {\cal L}=   \frac{1}{2} (\partial_{\mu}A_1)^2 - \frac{1}{2}(\partial_{\mu}A_0)^2 -  \partial^{\mu}  \bar{c} \partial_{\mu}c =
  \frac{1}{2} \partial_\mu \phi_2 \partial^\mu \phi_2 - \frac{1}{2} \partial_\mu \phi_1 \partial^\mu \phi_1- \partial^{\mu}  \bar{c}  \partial_{\mu}c
\ee    
    which is our original lagrangian (\ref{lag}) supplemented by the $c-$ ghost term. Selection of the physical Hilbert space is accomplished by considering the BRST closed states, i.e. the states which are annihilated by $Q_{BRST}$ operator.  This requirement   replaces the Gupta Bleuler condition (\ref{gb}) we used in the previous sections. 
    
    We proceed with the construction as follows. In addition to our expansion for $\phi_1, \phi_2$ fields, we also expand  the c-ghost field in Minkowski space in the same way, 
    \be
\label{c-expansion}
c (t, {x})&=&\sum_{k}\left[ c_k u_k(t,{x})+c_k^{\dagger}u_k^*(t,{x})\right] \, , ~~~~~ u_k(t,x) =\frac{1}{\sqrt{4\pi\omega}} e^{-i\omega t+ikx}, \\
\bar{c} (t, {x})&=&\sum_{k}\left[ \bar{c}_k^{\dagger} u_k^*(t,x)+\bar{c}_k u_k(t,{x})\right] \, , ~~~~~ \{c_k^{\dagger}, \bar{c}_{k'}\}=\delta_{kk'}, ~~ \{\bar{c}_k^{\dagger},  c_{k'}\}=\delta_{kk'}.  \nonumber
\ee
    One can construct the Minkowski space BRST operator as follows\footnote{ Our notations are different from ref. \cite{Evslin:2005zv}.
    Namely, we keep  our notations $a_k, b_k$ representing  $\phi_1, \phi_2$ fields (\ref{expansion}).  These operators enter the expansion  for  $A_0$ and $A_1$  fields    in notations (\ref{lag_BRST}). At the same time, in ref. \cite{Evslin:2005zv} $b_k$    describe   the temporal photon field $B=A_0$ while $a_k$   describe the combination $A_0+A_1$.}
    \be
    \label{BRST-M}
    Q_M=\sum_k \left[ c_k^{\dagger}b_k+c_k^{\dagger}a_k+a_k^{\dagger}c_k+b_k^{\dagger}c_k\right].
    \ee
     This operator obviously annihilates all physical states including the vacuum state,
    \be
    \label{Q-vacuum}
Q_M \left| {\cal H}_{\mathrm{phys}}\right> = 0, ~~~~~~ Q_M \left| 0\right> = 0
\ee
 The operator $Q_M$ can be written as an integral over a Cauchy surface of a local charge density $\rho_{M}$
    \be
    \label{rho}
    Q_M&=&\int dx\rho_M(x, t),\\ \nonumber 
    \rho_M&=&i\left[\phi_1(x, t)+\phi_2(x, t)\right]
\frac{\partial}{\partial t}c(x, t)-ic(x, t)\frac{\partial}{\partial t}\left[\phi_1(x, t)+\phi_2(x, t)\right].
    \ee
    
     Our next step is to construct the BRST charges for the Rindler observers which can be done in a similar way for $R$ wedge,
    \be
    \label{BRST-R}
    Q^R=\sum_k \left[ c_k^{ R\dagger}b_k^R+c_k^{R\dagger}a_k^R+a_k^{R\dagger}c_k^R+b_k^{R\dagger}c_k^R\right].
    \ee   
     and for $L$ wedge, 
   \be
    \label{BRST-L}
    Q^L=-\sum_k \left[ c_k^{ L\dagger}b_k^L+c_k^{L\dagger}a_k^L+a_k^{L\dagger}c_k^L+b_k^{L\dagger}c_k^L\right].
    \ee  
where sign $(-)$ is due to the fact that   time-like Killing vector $+\partial_{\eta}$ in R and  $-\partial_{\eta}$ in L, see eq. (\ref{R}), (\ref{L}).
In this expression the c-ghost fields in the Rindler space are defined in the same way as $\phi_1$ and $\phi_2$ fields, see eq. (\ref{expansion-R}), 
\be
\label{R-c}
c=\sum_k\frac{1}{\sqrt{4\pi\omega}}(c^L_ke^{ik\xi^L+i\omega \eta^L}+c^{L\dagger}_ke^{-ik\xi^L-i\omega \eta^L}+c^R_ke^{ik\xi^R-i\omega \eta^R}+c^{R\dagger}_ke^{-ik\xi^R+i\omega \eta^R}) \\ \nonumber
\bar{c}=\sum_k\frac{1}{\sqrt{4\pi\omega}}(\bar{c}^L_ke^{ik\xi^L+i\omega \eta^L}+\bar{c}^{L\dagger}_ke^{-ik\xi^L-i\omega \eta^L}+\bar{c}^R_ke^{ik\xi^R-i\omega \eta^R}+\bar{c}^{R\dagger}_ke^{-ik\xi^R+i\omega \eta^R}).
\ee
 These operators $Q^{(R,L)}$     annihilate their physical states including their corresponding vacuum states,
    \be
    \label{R-vacuum}
Q^{(R,L)} \left| {\cal H}_{\mathrm{phys}}^{(R,L)}\right> = 0, ~~~~~~ Q^{(R,L)} \left| 0^{(R,L)}\right> = 0.
\ee

Our next task is to compute $Q^R\left| 0\right>$.  This calculation will  tell us how   the Rindler observer moving with acceleration over Minkowski vacuum state $ \left| 0\right>$ makes the selection of the  physical states.   To perform the computations we have to express the BRST operator for the Rindler observer in terms of the combinations (\ref{analytic}) as we have done before in our previous computations for energy, see eq. (\ref{RR}). As we mentioned above,  the combinations (\ref{analytic})   are analytic, share the positivity frequency analyticity properties of the Minkowski modes (\ref{c-expansion}), and therefore,   share a common vacuum state $ \left| 0\right>$.
Therefore, we expand $c(x,t)$ field in the same way as we did for $\phi_1, \phi_2$ fields, see eq. (\ref{expansion-M1}), (\ref{expansion-M2}),
\be 
\label{c-R}
c(x,t)&=&\sum_k\frac{1}{\sqrt{ 4\pi\omega}}  \cdot \frac{1}{\sqrt{(e^{\pi\omega/a}-e^{-\pi\omega/a})}} \Big[c^1_k(e^{\frac{\pi\omega}{2a}+ik\xi^R-i\omega\eta^R}+e^{\frac{-\pi\omega}{2a}+ik\xi^L-i\omega\eta^L})  \nonumber \\
&+&c^2_k(e^{\frac{\pi\omega}{2a}+ik\xi^L+i\omega\eta^L}+e^{\frac{-\pi\omega}{2a}+ik\xi^R+i\omega\eta^R})  \nonumber \\
&+&c^{1\dagger}_k(e^{\frac{\pi\omega}{2a}-ik\xi^R+i\omega\eta^R}+e^{\frac{-\pi\omega}{2a}-ik\xi^L+i\omega\eta^L})\nonumber\\
&+&c^{2\dagger}_k(e^{\frac{\pi\omega}{2a}-ik\xi^L-i\omega\eta^L}+e^{\frac{-\pi\omega}{2a}-ik\xi^R-i\omega\eta^R})\Big],
\ee
where in addition to (\ref{vacuum-M}) the Minkowski vacuum state satisfies the following conditions formulated in terms of the basis (\ref{expansion-M1}), (\ref{expansion-M2}), (\ref{c-R}),
 \be
\label{c-vacuum-M}
a_k^1|0\>=0 \, , ~~~a_k^2|0\>=0 \, , ~~~ b_k^1|0\>=0 \, , ~~~  b_k^2|0\>=0 \, , ~~~ c_k^1|0\>=0 \, , ~~~c_k^2|0\>=0 \,   ~~~ \forall k \, .
\ee
The Bogolubov's coefficients can be computed in the same way as before (\ref{Bogolubov}). The result reads  
\be 
\label{Bogolubov-c}
c^L_k=\frac{e^{-\pi\omega/2a}c^{1\dagger}_{-k}+e^{\pi\omega/2a}c^2_k}{\sqrt{e^{\pi\omega/a}-e^{-\pi\omega/a}}}, ~~~~~~
c^R_k=\frac{e^{-\pi\omega/2a}c^{2\dagger}_{-k}+e^{\pi\omega/2a}c^1_k}{\sqrt{e^{\pi\omega/a}-e^{-\pi\omega/a}}}.  
 \ee
Using the Bogolubov's coefficients (\ref{Bogolubov}), (\ref{Bogolubov-c}) one can express the BRST operator (\ref{BRST-R}) for the R-Rindler observer in terms 
of basis (\ref{expansion-M1}), (\ref{expansion-M2}), (\ref{c-R}). The corresponding expression is quite long, and we do not really need it. What we actually need in order 
to demonstrate our main point, is the part of BRST operator $\Delta Q^R $  which contains exclusively  creation operators.     The corresponding part $\Delta Q^R $ can be represented as follows, 
\be
\label{delta_R}
\Delta Q^R= \sum_k\frac{1}{ (e^{\pi\omega/a}-e^{-\pi\omega/a})} \left[ c_k^{ 1\dagger}b_{-k}^{2\dagger}+b_k^{1\dagger}c_{-k}^{2\dagger}+c_k^{1\dagger}a_{-k}^{2\dagger}+a_k^{1\dagger}c_{-k}^{2\dagger}\right].
\ee
It is obvious that the BRST operator as defined by the Rindler observer does not annihilate the Minkowski vacuum as $Q^R$ has the  terms (\ref{delta_R}) 
  which do not annihilate the Minkowski vacuum, 
  \be
    \label{delta_R1}
  Q^{R} \left| 0 \right> =  \Delta Q^{R} \left| 0 \right> \neq 0. 
\ee
This conclusion is in accord  with  our previous result  (\ref{RR}) on computation of the energy $ \left< 0\right| H^R \left| 0 \right>\neq 0$ observed by the Rindler observer moving over Minkowski  vacuum.  The results (\ref{RR}) and (\ref{delta_R1})  are obviously consistent with each other\footnote{One can compute an additional term $\Delta\mathrm{H}^R_c $ to the hamiltonian due to  $c-$ ghost field.
For R-Rindler observer it takes the form $\Delta\mathrm{H}^R_c=\sum_k\omega_k(\bar{c}_k^{R\dagger}c_k^R-c_k^{R\dagger}\bar{c}^R_k)$ in addition to eq. (\ref{H-R}), where sign $(-)$ is resulted from anti-commutator  (\ref{c-expansion}) when normal ordering operation for  $\Delta\mathrm{H}^R_c$  is performed. Using the Bogolubov's coefficients (\ref{Bogolubov-c}) for c-fields one can explicitly check that two terms cancel each other  in vacuum expectation value $\left< 0 \right| \Delta\mathrm{H}^R_c\left| 0 \right>=0$  such that there is no net contribution to energy from $c-$ field, as expected. While the structure of  the hamiltonians  for $\phi_1, \phi_2$ fields  (\ref{H-R}) and $\Delta\mathrm{H}^R_c$ along with  the corresponding Bogolubov's coefficients are similar,  the contributions to the energy are different. Technically, it is due to the fact that the commutation relations for 
$a_k$ and $b_k$ operators have opposite signs, see eqs. (\ref{comm_2}), (\ref{comm_1}), while for $c_k, \bar{c}_k$ operators    the    signs are the same, see eq.(\ref{c-expansion}). Non-technical, intuitive explanation for this effect will be given in the discussion section \ref{results}, and essentially related to the fact that $\phi_1$ is sensitive to the topological sectors of the theory as the topological density operator 
$\epsilon_{\mu\nu}F^{\mu\nu}$ is expressed in terms of $\phi_1$ ghost field, while the Fadeev Popov ghosts are design to cancel the 
unphysical polarizations of the gauge fields in the bulk of the space-time, and not sensitive to the boundary/horizon effects, and therefore, can not be sensitive to the acceleration parameter $``a"$.} , and show that extra energy  (\ref{RR})
is resulted from the states which carry  non-vanishing BRST charge (\ref{delta_R1}).

In BRST approach one can explicitly see how 
the cancellation for the Minkowski BRST operator actually works. To see this we need the expression for the BRST operator $Q^L$ for the L- Rindler 
observer along with $Q^R$. More precisely, we need its $\Delta Q^L$ part  containing  the   creation operators only.  It is given  by
 \be
\label{delta_L}
\Delta Q^L= -\sum_k\frac{1}{ (e^{\pi\omega/a}-e^{-\pi\omega/a})} \left[ c_k^{ 2\dagger}b_{-k}^{1\dagger}+b_k^{2\dagger}c_{-k}^{1\dagger}+c_k^{2\dagger}a_{-k}^{1\dagger}+a_k^{2\dagger}c_{-k}^{1\dagger}\right],
\ee
 where sign $(-)$ is due to the fact that   time-like Killing vector $+\partial_{\eta}$ in R and  $-\partial_{\eta}$ in L, see eq. (\ref{BRST-L}). 
 The crucial observation is that appropriate Minkowski BRST charge expressed in basis (\ref{expansion-M1}), (\ref{expansion-M2}), (\ref{c-R})
 is the combination of two, $Q^M=Q^R+Q^L$, see \cite{Evslin:2005zv}, such that the dangerous terms $\Delta Q^L$ and $ \Delta Q^R $   are exactly cancelled $\Delta Q^L+ \Delta Q^R =0$. The cancellation  (between positive $k$ from $ \Delta Q^R $ and negative $k$ from $ \Delta Q^L $) can be explicitly seen from (\ref{delta_R}), (\ref{delta_L}) where summation over entire $k$ interval is assumed, 
 $k\in (-\infty, +\infty)$. 
 
 The most important lesson from this cancellation can be formulated  as follows. The BRST operator as constructed by the Rindler observer does not annihilate the Minkowksi vacuum state because $Q^R$ and $Q^L$ are integrals of BRST charge density over half of space and both contain terms of the form $c_k^{ 2\dagger}b_{-k}^{1\dagger}$ which do not annihilate Minkowski $ \left| 0 \right>$. At the same time an inertial observer in  $ \left| 0 \right>$ observes a universe with no net BRST charge as a result of cancellation between $Q^L$ and $Q^R$ while a  BRST charge density (\ref{rho}) does  not vanish separately   in $L$ and $R$ parts.   The result of such cancellation as seen by Minkowski observer  is  in drastic contrast with   measurements performed by the Rindler  observers  who do not ever have access to the entire space time. Therefore, from the Rindler's view point the cancellation in Minkowski space can be only achieved if one uses both sets (L and R). Of course, using the both sets would contradict to  the basic principles as the R-Rindler observer does not have access to the L wedge even  for arbitrary small  acceleration parameter $a$. Therefore, a Rindler observer with access to only part of the universe will see a net BRST charge as eq. (\ref{delta_R1}) states.

 One should emphasize that this  effect manifests itself only globally, not locally. Indeed, if we 
define a  subspace of physical states $ |{\cal H}_{\mathrm{phys}}\rangle_{\bar{k}}$ where all modes have momenta $k> \bar{k}\gg a$  such that the notion ``particles"  becomes   well defined for this subspace, than   the deviation from the standard local physics   will   be    strongly    suppressed as one can see from eq. (\ref{delta_R}) where  $Q^{R} |{\cal H}_{\mathrm{phys}}\rangle_{\bar{k}}  \sim \exp(-\frac{\pi\bar{k}}{a})$.  In cosmological context when $a$ is identified with $H$ as we already mentioned this suppression  is astonishingly small $\sim \exp(-\frac{\bar{k}}{H})\sim \exp(-10^{27})$.  With this accuracy  $Q^{R} |{\cal H}_{\mathrm{phys}}\rangle_{\bar{k}}  =0+{\cal{O}} ( \exp(-10^{27}))$ which is indistinguishable  from Minkowski space result (\ref{Q-vacuum}). The only modes which will be  affected  
are those with the  wave lengths of  order   $\lambda\sim k^{-1}\sim a^{-1}$ when the entire notion of ``particle" is not even defined.

    \section{Interpretation. Speculations. Concluding remarks. }
   First,   we conclude with the main results of our studies. We follow with   interpretation of   these results  by presenting   some analogies from condensed matter physics. Finally, we comment on observational consequences of the obtained  results. 
   \subsection{Basic Results.}\label{results}
        $\bullet$ Exactly solvable model considered in this work  (2d Maxwell system defined on the Rindler space)   supports the picture advocated  in  \cite{UZ} that there will be an extra contribution  to the vacuum energy  in a nontrivial  background   in comparison with Minkowski space time.
    This extra contribution can be  traced to the massless topological degrees of freedom which belonged  to unphysical Hilbert space (in Minkowski space). \\
    $\bullet$ The technical reason for this effect to occur  is the property of Bogolubov's coefficients which mix the positive and negative frequencies modes. The corresponding  mixture can not be avoided because  the projections to 
     positive -frequency modes 
  with respect to Minkowski time $t$  and  positive -frequency modes    with respect to the Rindler observer's proper  time $\eta$  are  not equivalent. 
  The exact cancellation of unphysical degrees of freedom which is maintained in Minkowski space can not hold in the Rindler space.\\
   $\bullet$ In BRST approach this effect manifests itself as the presence of BRST charge density    in $L$ and $R$ parts.   An inertial observer in  $ \left| 0 \right>$ observes a universe with no net BRST charge only as a result of cancellation between the two.  However,  the Rindler  observers  who do not ever have access to the entire space time
   would see a net BRST charge. Therefore, they operate with the states which do not belong to the physical subspace of the  inertial observer in  Minkowski space $ \left| 0 \right>$.\\
   $\bullet$ We emphasize once again: the effect under study is  exclusively due  to the nontrivial topological sectors of   gauge theory. The  effect may only occur in theories like 2d QED or 4d QCD with  nontrivial topological structure 
   of the ground state\footnote{ One should remark here  that the topological structure also  emerges for accelerating 4d QED, as there are only two nontrivial coordinates in this system: time $t$ and the direction of acceleration $x$ while $yz$ plane is decoupled from the accelerating system and can be ignored for the present studies. Therefore, at very large distances (very low energies)  the model becomes effectively  a 2d  theory when a nontrivial topological structure of the theory emerges. }.     There are no any extra propagating degrees of freedom in our framework as explained in next section \ref{interpretation} below.   In particular,  Kogut Susskind  ghost represented by $\phi_1$ field saturates the contact term in the topological susceptibility, and therefore it  effectively accounts for the 
  summation over all topological sectors of the theory, see Appendix C for details.   All other types of ghosts, including the conventional Fadeev Popov ghosts   are not related to topologically nontrivial sectors of the theory. They are design to cancel  the contributions of unphysical polarizations of the gauge fields in the bulk of the space-time. This  conventional  cancellation is expected to hold in all cases,    including  the accelerating frame and arbitrary  curved background. In this respect, the Veneziano ghost in 4d (Kogut Susskind ghost in 2d) are unique: due to their topological nature and their direct connection  to the topological density operator ($\epsilon_{\mu\nu}F^{\mu\nu}$ in 2d and $F_{\mu\nu}^a\widetilde{F}^{a\mu\nu}$ in 4d  are explicitly expressible   in terms of the $\phi_1$ ghost field) they describe physics at the boundaries/horizons, and therefore they are very sensitive to the global characteristics of the entire space-time.

  \subsection{Interpretation.  }
  \label{interpretation}
  As explained in length in the text the nature of the effect (extra amount of the vacuum energy   observed by the Rindler observer in comparison with the Minkowski observer) is    the same as  the  conventional Unruh effect\cite{Unruh:1976db} when the  Minkowski  vacuum $ \left| 0 \right>$ is restricted to the Rindler wedge  with no   access to the entire space time.  Precisely the same restrictions lead to a non vanishing BRST charge density   in $L$ and $R$ parts taken separately while it vanishes for the entire Minkowski space.  This result, by definition, implies that
   the states which were unphysical (in Minkowski space) lead to physically observable phenomena, though it can not be interpreted in terms of pure states of individual particles, see below.  
The effect is obviously sensitive to the presence of the horizon  and/or the boundaries, and, therefore is  infrared in nature. An appropriate description in this case, as is known,    should be  formulated (for $R$ observer) in terms of the density matrix by ``tracing out" over the degrees of freedom associated with $L$-region.    This procedure  leads, as is known,  to some     correlations between causally disconnected regions of space-time,  though  those correlations can not be used to send signals\cite{Unruh:1976db},\cite{Unruh:1983ms}, see Appendix A for the details.
  
  Is it a real physical effect? One should remind the reader that a concern of the ``reality"  of the Unruh radiation was  unsettled until the paper \cite{Unruh:1983ms} appeared, see also~\cite{Birrell:1982ix}. The paper was 
  specifically devoted to the ``reality" issue. To be more specific, the authors of ref.  \cite{Unruh:1983ms} consider a simple particle detector model to demonstrate that 
  the radiation is a real physical phenomenon resolving  a number of paradoxes related to causality and energy conservation. An important for the present work result of ref.  \cite{Unruh:1983ms} is as follows: the absorption of a Rindler particle corresponds to emission of a Minkowski particle without violation causality and energy conservation. Now we want to repeat a similar analysis to see if any physical  radiation really occurs in our case when the  system is described by two fields $\phi_1, \phi_2$ with opposite commutation relations (\ref{lag}), (\ref{comm2}), (\ref{comm1})  instead of a single physical massless field in ref. \cite{Unruh:1983ms}. The crucial observation for future analysis is as follows:
  the fields $\phi_1, \phi_2$ which are originated from unphysical (in Minkowski space) degrees of freedom can couple to  other fields only through a combination $(\phi_1-\phi_2)$ as a consequence  of the original gauge invariance. In particular, it has been explicitly demonstrated  in 2d QED~\cite{KS} and in 4d QCD~\cite{UZ} where the corresponding 
  interaction to the physical degree of freedom  $\hat\phi $ takes the form $ \sim \cos\left[\hat\phi + \phi_2 - \phi_1 \right]$. Precisely this property along with Gubta-Bleuler auxiliary condition (\ref{gb}), (\ref{gb1}) provides the decoupling of physical degrees of freedom from unphysical combination $(\phi_2 - \phi_1)$
  as discussed in great details in \cite{KS}. 
  
  Now, in order  to repeat analysis of ref.  \cite{Unruh:1983ms} we have to replace a single physical field $\Phi$ from ref. \cite{Unruh:1983ms} by specific combination $(\phi_2 - \phi_1)$ fields for our system (\ref{lag}).  It leads to some drastic consequences as instead of conventional expectation values such as 
  $<0| a_k ... a^{\dagger}_{k'}|0>\neq 0$ from ref. \cite{Unruh:1983ms} we would get $<0| (a_k-b_k)... (a^{\dagger}_{k'}-b^{\dagger}_{k'})|0>= 0$. The corresponding  matrix elements vanish as a result   of the commutation relation $[(a^{\dagger}_{k'}-b^{\dagger}_{k'}), (a_k-b_k)]=0$ which follows from (\ref{comm2}), (\ref{comm1}). Furthermore,  as $[\mathrm{H}, (a_k-b_k)]=(a_k-b_k)$ the structure 
  $(a_k-b_k)$ is preserved such that $a_k$ and $b_k$ never appear separately. Based on this observation, 
    one can argue that the same property holds  for any other operators which constructed from the combination $(\phi_2 - \phi_1)$. In different words, no actual radiation of real particle occurs in our case in contrast with real Unruh radiation given by formula (3.29) from ref. \cite{Unruh:1983ms}.  The same conclusion  also follows from analysis of  the
    Wightman Green function describing the dynamics of the field, see Appendix B for details. Therefore, there is an extra energy 
  in the system observed by a Rindler observer (\ref{RR}) without radiation of any real particles.  In many respects, this feature  is similar to the Casimir energy though spectral density distribution (\ref{RR}) describing the fluctuations of the vacuum energy has a nontrivial $\omega$ dependence in contrast with what happens in the Casimir effect.
  
$\bullet$   Based on the comments presented above, we interpret the extra contribution to the energy observed by the Rindler observer as  a result of  formation of a specific   configuration   which can be coined as  the ``ghost condensate" (similar to the QCD gluon condensate  which  effectively accounts  the  physics in the infrared, $k\leq \Lambda_{QCD}$) rather than a presence of ``free particles"   prepared in a specific mixed state\footnote{The corresponding  spectral density distribution  saturating this ghost condensate in our simple  2d model is determined by eq. (\ref{RR}), see also Appendix A
on construction of the density matrix. However the spectral density distribution  would be quite different  in a more realistic case of FLRW universe when $a$ (which plays the role of a Hubble constant $H$) effectively becomes a time-dependent  parameter and the interaction  is not neglected. 
}. In different words, we interpret the ghost contribution to the energy as a convenient way to account for a nontrivial  infrared physics at the horizon  and/or  the boundary.  It is  possible that the same physics, in principle is describable  without the ghosts  (which are typically  introduced as auxiliary fields to resolve constraints
and  avoid nonlocal expressions in a hamiltonian), see Appendix C for the details.   
 However it is quite  likely that such a description would be much more (technically) complicated in comparison with 
the presented technique as it would deal with singularities and regularization problems  which always accompany horizon/boundary regions.

Let us present an additional argument supporting this interpretation. 
Let us  assume that in the remote past and future the space-time is Minkowskian one while in the middle we have a situation where 
the positive and negative frequency modes mix which resulting nonzero contribution to the energy from unphysical (in Minkowski space) modes. In this case  in the remote past and future the notion of particle is well defined.  In fact, there is a simple 2d model with a specific profile for the expansion function $a(t)$ interpolating between two Minkowski spacetimes which can be solved exactly.  The outcome (see sec.~3.4 in~\cite{Birrell:1982ix}) is that, even in this plain example $\beta_{kl}\neq0$, which can be understood as a production of particles by the expanding background. In our case this should not be interpreted as actual emission of ghost modes, as the ghost modes are not the asymptotic states in Minkowski spacetime  in the remote past and future, and therefore they can not propagate to infinity in contrast with conventional analysis~\cite{Birrell:1982ix}. Rather, one should interpret   (\ref{H2}) in general and (\ref{RR}) in particular for the Rindler space, as  an additional time dependent contribution to the vacuum energy in time dependent background in comparison with Minkowski space-time. 

This extra energy is  entirely ascribable to the presence of the unphysical (in Minkowski space) degrees of freedom.  However, we can not interpret them as being particles in the intermediate region where entire notion of particle is not well defined~\cite{Birrell:1982ix}, and also, we can not detect them in the remote past and future as they are not a part of physical Hilbert space. Moreover, from eq. (\ref{RR}) we can not specify the localization of this extra energy  as the typical wavelengths of the fluctuations are order of the horizon scale.
Therefore, we interpret this contribution to the energy in the intermediate region as a result of a time-dependent ``ghost condensation" of pairs   with opposite momenta, when creation operators from different causally  disconnected regions $L$ and $R$ 
 enter the relevant expression for the vacuum state, see eq. (\ref{squeezed}) in Appendix A for precise definition.
While  one can not answer the question about the localization of this energy,   one can argue, using a different approach  that the energy (as well as the entropy associated with it)  is actually localized  exactly at the horizon, see below. This extra energy interacts with the gravity field as it can be measured by the Rindler particle detector, and passes all tests to be identified with the dark energy as argued in \cite{UZ}, see also few comments on this below in sections~ \ref{comments} and \ref{observations}.

As we mentioned earlier, this is not the first time when unphysical (in Minkowski space) ghost
contributes to a physically observable quantity.   The first example is   the famous resolution of the $U(1)_A$ problem in QCD\footnote{See also another approach due to Witten\cite{witten} where the ghost does not even appear in the system. However, the corresponding physics due to a nontrivial   background
 does not go away in the Witten's formulation, see few  comments on the   Witten's approach  in a curved background in ref.\cite{UZ}. }  by  Veneziano~\cite{ven}, when the Veneziano ghost  being  unphysical   nevertheless provides a crucial contribution into  the  gauge invariant correlation
function (topological susceptibility). 

\subsection{Few More Comments.}\label{comments}
  The next   comment we want to make  can be formulated   as follows. Our starting point was lagrangian (\ref{lag}) which describes QED in the Lorentz covariant  gauge.
 Instead, we could choose a Lorentz non-covariant gauge, for example the  Coulomb gauge, such that $\phi_1$ and $\phi_2$ fields  would not even appear
 in the system, as the introduction of these auxiliary  fields is essentially only a matter of convenience (helping to resolve constraints and non-localities).  Where does effect go in these gauges? The point is that the description in the Coulomb and similar gauges (when formally  only the physical degrees of freedom 
 remain in the system) leads to an extra  term in the lagrangian which  is completely determined by the boundary conditions, and which is normally ignored in description of local physics. This term, in particular,  is related to the  classification of   the allowed large gauge transformations with nontrivial topological conditions at the boundary. These   features of pure gauge, but still topologically   nontrivial configurations, eventually lead to the construction of the so-called $\left|\theta\right>$ vacuum state which represents an infinite series of degenerate the so-called ``winding states", see e.g.\cite{KS}.  
 We advocate the ghost- based technique  to account for this physics because the corresponding   description   can be easily generalized into curved background, while a similar  generalization  (without the ghost, but with explicit 
accounting for the infrared behaviour at the boundaries) is unknown and  likely to be much more technically complicated. In different words, the description in terms of the ghost is a matter of convenience which allows us to  account  for  the boundary effects  in topologically nontrivial sectors of the theory.

The relation between the two approaches can be explicitly worked out in a simple 2d model, see Appendix C for the details.
The example from Appendix C shows, in particular,  that  even when  there are no physical photons in the system,  still there is an extra term sensitive to the boundaries
and large distance physics.  Therefore, our claim\cite{UZ} that there is a  mismatch between the vacuum energy computed in slowly expanding universe   and the one which is computed in flat Minkowski space  should not be very surprising after all: in both cases there is a sensitivity  to the boundary conditions (which   are very different   in these two cases). 

From our discussions in section 3 it should be  quite obvious that the corresponding term  for the Rindler space and for Minkowski space would be different  because the allowed large gauge transformations  in  Minkowski space   and in the Rindler space   are  not equivalent. However, an explicit construction is still lacking as it would require an infrared regularization (e.g. similar to the one used in  Appendix C for Minkowksi space) to classify the large gauge transformations. Presently we do not know how to do it consistently in Minkowski and Rindler spaces.  Another benefit of  dealing with the extra ghost degrees of freedom   
 is  the possibility to avoid all difficult questions on imposing some nontrivial consistent boundary conditions at the horizon and/or the boundaries when a singular behaviour is unavoidable.
 
Our next comment is as follows. 
 The interpretation of the effect  in terms of BRST charge suggests an analogy with some condensed matter systems.  To be more precise, consider the so-called charge fractionalization  effect in a system  which admits solitons, see\cite{Niemi:1984vz} for review. The effect in few  lines can be explained as follows.  In the soliton sector of the theory due to the presence of a single zero fermion mode the soliton esquires a fermion charge $1/2$ as a result of the double degeneracy in the soliton sector of the theory. The charge is localized in the region which is order of   a   soliton size $l$. Original underlying theory was defined with integer charges only. Therefore, the question is: where does another $-1/2$ go? The answer is: it goes to the boundary of a sample with arbitrary large size $L$ such that an experimentalist-R   with no access  to the scales of order $L$ would see charge $1/2$.  At the same time, an experimentalist-M with access
 to the entire sample including the boundaries, would measure the total charge 0. This picture resembles our system in a number of aspects when  
 experimentalist-R is analogous to the Rindler observer while experimentalist-M plays the role of Minkowski inertial observer.  A fractional charge observed by 
 experimentalist-R is analogous to a non vanishing BRST charge measured by a Rindler observer (\ref{delta_R1}), while a vanishing total charge measured by   Minkowski inertial observer is analogous to BRST charge $Q^M$.  The role of the boundary $L$ of a sample is analogous to the horizon scale. The charge fractionalization  effect in condensed matter physics is obviously has infrared  nature though it is often derived by using a technique which requires  a  summing up an arbitrary high frequency  modes, see\cite{Niemi:1984vz} for details. 
 
 Our final comment is on relation between two different frameworks: first is based on the hamiltonian approach advocated in this work, while the 
 second approach is based  on computation of the renormalized stress tensor $\la T_{\mu\nu}\ra_{\text{ren}}$. One could naively think that using the conventional
 transformation law (by transforming $T_{\mu\nu}$ from Minkowski to the Rindler space) one should always get the vanishing result for
 the renormalized  stress tensor $T_{\mu\nu}$ even for an accelerating observer (moving over Minkowski spacetime)
    performing the measurements using his particle detector. It is known why this argument in general  is not correct, see explicit computations in refs~\cite{Unruh:1992sw, Sanchez:1985ys, Parentani:1993yz}. The key point lies
 \exclude{ in the normal ordering in Rindler space-time 
 in the course of computation of the relevant Green's function. The corresponding normal orderings  differ in different space-times,
 leading to }
 in a complicated subtraction procedure in the corresponding Green's function which itself is extremely singular object at coinciding points and requires special  care in subtractions. In particular one can indeed demonstrate 
  that $ \la T_{\mu\nu}\ra_{\text{ren}} = 0$   in the bulk of the space-time as a result of cancellation of two singular expressions~\cite{Unruh:1992sw}. However, the Rindler  particle detector would measure a non-vanishing $ \la T_{\mu\nu}\ra_{\text{ren}} \neq 0$  exactly on the horizon where two singularities collapse~\cite{Unruh:1992sw}. 
  One should also remark here that the expression for the Planck spectrum (\ref{RR})  does not specify the coordinate localization of the energy and entropy as measured by the Rindler observer. Computation of $ \la T_{\mu\nu}\ra_{\text{ren}} \neq 0$ on the horizon answers the question regarding the localization. Therefore, one should treat both approaches as complimentary to each other. One should also add that vanishing $ \la T_{\mu\nu}\ra_{\text{ren}} = 0$ in the bulk 
  is a result of very nontrivial cancellations  between emission and absorption of the energy with involvement of an  external accelerating agent. The accelerating agent  is not a part of the system studied in this paper as we do not accelerate the system, but assume that a constant  acceleration is   produced by some external forces. This is   the source of  vanishing $ \la T_{\mu\nu}\ra_{\text{ren}} = 0$ in the bulk  of space-time as an accelerating agent 
  is not part of the system, see original papers on the interpretation of the effect, \cite{Unruh:1976db,Unruh:1983ms,Unruh:1992sw}.  
  
   The  relation between these two different frameworks  shows once again that all nontrivial effects considered in  this paper    are due to the behaviour of the system in far- infrared, on the horizon separating two sub-systems.      The Planck spectrum    emerges as a result  of the description    in terms of the density matrix in $R$ region by ``tracing out" over the degrees of freedom associated with 
 inaccessible states in $L$-region. In different words, 
  the Bogolubov transformations  
  describe a  construction  when a total system is divided into two subsystems with the horizon separating them.  This is the deep physics reason why  the Planck spectrum  emerges for a subsystem. 
 It is known that a number of nontrivial physics effects (including the entanglement)
  are described by the common surface separating such two sub- systems.
  Therefore,   the Bogolubov transformations is not a trivial change of basis (as one could naively think), 
but in fact an appropriate tool to describe a very deep property of entanglement. 
   The corresponding physics is described by the common surface separating such two sub- systems, i.e. by the horizon separating $L$ and $R$ Rindler wedges.   In different words,  a non-vanishing $ \la T_{\mu\nu}\ra_{\text{ren}} \neq 0$  exactly on the horizon  as measured by  the Rindler  particle detector is in complete accordance with our interpretation of the effect presented above.

  \subsection{Observational  consequences}\label{observations} 
 
 $\bullet$ The obtained results may have some   profound consequences for our understanding of physics at the largest possible scales in our universe.  First of all,  the dark energy observed in our universe might be  a result of mismatch between the vacuum energy computed in slowly expanding universe with the expansion rate $H$  and the one which is computed in flat Minkowski space\cite{UZ}. If true, the difference between two metrics would lead to an estimate $\Delta E_{vac}\sim H\Lambda_{QCD}^3\sim (10^{-3} {\text eV})^4$ which is amazingly close to the observed value today. The process of energy pumping   will continue as long as our space-time  is deviated from flat Minkowski  metric. This extra energy interacts with the gravity field, and passes all tests to be identified with the dark energy as argued in \cite{UZ}, see also some comments below. 
 The fate of our universe in this paradigm is determined (eventually) by the feedback reaction on the gravity field. This subject is   beyond the scope of the present work, and has not been  discussed here.  
 
 The most important feature of this  mechanism is the spectrum of the  fluctuations:  the typical wavelengths $\lambda_k$ of excitations  associated with  energy   (\ref{RR}) are of the order of the inverse Hubble parameter, $\lambda_k\sim 1/k\sim 1/H\sim 10\textrm{~Gyr}$.  Therefore, these modes  do not clump on distances smaller than $H^{-1}$, in {\it contrast with all other types of matter}, and therefore, this type of energy  passes  the crucial test  allowing it to  be identified with the observed dark energy.

At the same time, the localization pattern of such energy   in FLRW expanding universe and in our toy model (described by the Rindler metric)  differ. Indeed,  in FLRW expanding universe,  $ \la T_{\mu\nu}\ra_{\text{ren}}\neq 0 $ everywhere,  and the horizon  (where the effect  is  localized) changes its position/size with time  by  slowly filling the bulk of entire  space with time dependent fluctuations of wavelengths
 $\lambda_k (t)\sim 1/H (t)$ during the expansion of the  universe. Also, the source of the expansion is the part of the system which pumps the energy into    the formation of these long wavelengths  topological fluctuations.  
At the same time, in our toy model with the Rindler metric, the position of the horizon where $ \la T_{\mu\nu}\ra_{\text{ren}}\neq 0 $ as measured by the Rindler particle detector  is fixed once and forever. Moreover, the accelerating agent which is the source of the energy in our toy model is not a part of the system. 
In this respect, the relation between our study in the Rindler metric and 
FLRW expanding universe is analogous  to the Unruh effect (where  $ \la T_{\mu\nu}\ra_{\text{ren}} = 0$ in the bulk)  and the Hawking radiation from a black hole.

  $\bullet$ Furthermore, the same topological (unphysical in Minkowski space) degrees of freedom which is the subject of the present work may in fact lead to the Casimir type effect  as   argued  in~\cite{our4d} when   no massless propagating physical degrees of freedom are present in Minkowski space. 
  This effect can be exactly computed in a toy 2d QED model\cite{toy} which  is known to be a system with a single massive degree of freedom
  when massless unphysical degrees of freedom are decoupled in Minkowski space.
  Still, the Casimir like effect is  present in this 2d system~ \cite{toy}. The Casimir type effects 
  in 4d QCD also  appear to be present  
  on the lattice where the power like behaviour $(1/L)^{\alpha}$ as a function of the total   lattice size $L$  has been observed in measurements of the topological susceptibility~\cite{Gubarev:2005jm}. Such a behaviour is in huge contrast with  exponential $\exp(-L)$ decay  law which   one normally expects for any  theories with  massive degrees of freedom\footnote{ I am thankful to Misha Polikarpov who brought the paper \cite{Gubarev:2005jm} to my attention.}.
  
 $\bullet$ Also, it has been   argued in~\cite{cmbt} that  these effects  at very large scales could in principle be tested in upcoming CMB maps (PLANCK),
 including P-parity violating effects at very large scales.
 
 $\bullet$ Furthemore, a nature of the   magnetic field with characteristic intensity of around a few $\mu$G   correlated on very large scales and observed today  is still unknown\footnote{Originally, large-scale magnetic fields have been first discovered in our Milky Way  with $\mu$G intensity. Later on the magnetic fields of very similar strengths have been observed in clusters of galaxies, where they appear to be correlated over larger distances reaching the Mpc region. It is important to notice that such fields are not associated with individual galaxies, as they are observed in the intergalactic medium as well. 
  Finally, the most recent observations hint towards a possible magnetization of gigantic supercluster structures pushing the correlation lengths further away up to fractions of Gpc.}.
  One can argue that the very same (unphysical in Minkowski space)  degrees of freedom which is the subject of the present work may in fact induce
 the large scale magnetic field as a result of anomalous interaction with photons~\cite{Urban:2009sw}. More than that,  the corresponding induced magnetic field would naturally have the  intensity  $ B \simeq \frac{\alpha}{2\pi} \sqrt{H \Lqcd^3}\sim$ nG, which  by simple adiabatic compression during the structure formation epoch, could explain the field observed today at all scales, from galaxies to superclusters~\cite{Urban:2009sw}.
 
  $\bullet$ Finally, hadron production studies in a variety of high energy collision experiments have shown a remarkably universal feature, indicating a universal hadronization temperature $T\sim (150- 200) ~{\text MeV}$.
  From $e^+e^-$  annihilation to $pp$ and $p\bar{p}$  interactions and further to collisions of heavy nuclei, with energies from a few GeV up to the TeV range, the production pattern always shows striking thermal aspects, connected to an apparently quite universal temperature around $T\sim (150- 200) ~{\text MeV}$. 
  Such a thermal spectrum is observed even in cases when   conventional ``kinetic thermal equilibrium" can never be reached. 
  We argue in \cite{Zhitnitsky:2010zx}
  that this apparent thermalization can be understood as   a manifestation  of the Unruh effect
 through the event horizon, which itself dynamically emerges as a result of the confinement in the strongly interacting gauge theory.  We also argue that the violation of local ${\cal P}$ and ${\cal CP}$ invariance in QCD, as it is observed at RHIC, Brookhaven, is the direct consequence  of the ghost fluctuations with $0^{+-}$ quantum numbers considered in this paper. All these effects   occur as a result 
 of  restriction of the Minkowski vacuum $ |0\>$ to the Rindler wedge with no access to the entire space time (\ref{squeezed}), which is the key element  in all discussions of the property of entanglement.

 Finally, one should  note that QED photons, including unphysical polarizations, may also in principle contribute to dark energy.  This contribution however is very small, as it is of order of $L^{-4}$ or $H^4$ by dimensional reasons (see however the mechanism proposed in~\cite{qed1,qed2}).

 \section{Acknowledgements}
 I  would like to   thank  Gordon Semenoff who  suggested to test the ideas formulated in \cite{UZ} using the Rindler metric. 
 I also thank Gordon for the  discussions and collaboration  during  the initial stage of this project. I would like to  thank Bill Unruh for the discussions. 
 I am also  thankful to  Larry McLerran,  Valery Rubakov, Sergei Sibiryakov and  Arkady Vainshtein  and other participants of the   workshops at Brookhaven, April 26-30, 2010 and Cargese, May 10-15, 2010 where this work has been presented, 
  for useful and stimulating discussions. 
 This research was supported in part by the Natural Sciences and Engineering Research Council of Canada.

 \appendix
 \section{Squeezed state}
 The main goal of this Appendix is to construct the so-called squeezed state. We  also make few 
  comment  on the   correlations between causally disconnected regions of space-time which follow from this construction. 
  
 The explicit expression for the Bogolubov's coefficients (\ref{Bogolubov}) between Minkowski and Rindler spaces allows us to construct explicitly the so-called  ``squeezed state" which relates Minkowski and the Rindler vacuum states.
   The corresponding relation reads:
 \be
 \label{squeezed}
 |0\>= \prod_k\frac{1}{ \sqrt{(1-e^{-2\pi\omega/a})}}\exp\left[   e^{-\pi\omega/a}\left(b_k^{R\dagger}b_{-k}^{L\dagger}-
 a_{-k}^{R\dagger}a_{k}^{L\dagger}\right)\right]   \left| 0^{R}\right>  \otimes \left| 0^{L}\right>, 
 \ee
 where we take into account that the operators in the $L, R$ basis correspond to the decompositions with support in only one wedge such that the right hand side is represented by the tensor product $ \left| 0^{R}\right>  \otimes \left| 0^{L}\right>$.
 This relation is almost identical to the construction discussed in refs.\cite{Unruh:1976db} and  \cite{Unruh:1983ms},
 when the operators from  different causally disconnected regions $L$ and $R$ enter the same expression. The only difference is that two different types of operators $a_k$ and $b_k$ enter expression (\ref{squeezed}) corresponding to the ghost and its partner. 
 The relative sign minus in eq. (\ref{squeezed}) is due to the different signs in commutation relations (\ref{comm_2}) and (\ref{comm_1}) describing $\phi_1$ and $\phi_2$ fields. 
 As discussed in refs.\cite{Unruh:1976db} and  \cite{Unruh:1983ms}, 
 one can not use the correlations explicitly present in eq. (\ref{squeezed}) in order to send signals. 
 
 The expression (\ref{squeezed})  
 (while formally  similar) nevertheless is very different from analogous   formula for the corresponding ``squeezed state"
for conventional cosmological particle production. In our case the combination $a_k^{R\dagger}a_{-k}^{L\dagger} $ 
(with  operators from  different causally disconnected regions $L$ and $R$)
enters the expression (\ref{squeezed}) while in a   case of particle production one and the same   operator $a_k^{\dagger}$  appears twice in   combination  $\sim a_k^{\dagger}a_{-k}^{\dagger} $     entering the  relevant formula. 

 Finally, we should note that the Minkowski vacuum $|0\>$ is a pure state, but  it becomes the mixed state when restricted to a single Rindler region. One can construct the corresponding density matrix for $R$ region by ``tracing out" over the degrees of freedom associated with $L$ region exactly as it has been done  in refs.\cite{Unruh:1976db} and  \cite{Unruh:1983ms}.
 We shall not elaborate on this issue in the present paper. Rather, we want to emphasize once again that the basic reason
 for nonzero contribution to the vacuum energy in our case (\ref{RR}) is exactly the same as for the conventional Unruh effect. Namely,  it is due to the restriction of the Minkowski vacuum $ |0\>$ to the Rindler wedge with no access to the entire space time. The interpretations for the two cases however  differ: we interpret an additional energy as the ``ghost condensate" 
 of pairs $ a_{-k}^{R\dagger}a_{k}^{L\dagger}$   and $b_k^{R\dagger}b_{-k}^{L\dagger}$ in different causally disconnected regions $L$ and $R$ with opposite momenta, 
  rather than a presence of Òfree particlesÓ prepared in a specific mixed state defined by the temperature $T=\frac{a}{2\pi}$ (which is  the conventional interpretation for the Unruh effect). The main reason for these
   differences in  interpretation is discussed
  in Section 5.2 and Appendix B, and we refer the reader to the corresponding subsections for details.
  
   \section{Particle detector for the ghost. }
 As is known the ``reality" issue discussed in section \ref{interpretation}
can be formulated by considering the particle detector moving along the world line described by some function $x^{\mu}(\tau)$ where $\tau$ is the detector's proper time. In the case for the Rindler space the corresponding $\tau$ is identified with $\eta$ defined by formula (\ref{eta}). As is known, the corresponding analysis
in the lowest order approximation   is reduced to study of the positive frequency Wightman Green function defined as 
\be
\label{wightman}
D^+ (x, x')= \< 0|\Phi (x), \Phi (x') |0\> , 
\ee
while the transition probability per unit proper time is proportional to
its Fourier transform, 
\be
\label{fourier}
\sim \int^{+\infty}_{-\infty} d(\Delta \tau)e^{-i\omega\Delta \tau} D^+ (\Delta\tau) 
\ee
where we use notations from ~\cite{Birrell:1982ix}.   
 In case of inertial trajectory for massless scalar field $\Phi$  the positive frequency Wightman Green function
 is given by 
 \be
D^+ (\Delta\tau')= -\frac{1}{4\pi^2} \frac{1}{(\Delta\tau -i\epsilon)^2}
\ee
and the corresponding Fourier transform (\ref{fourier}) obviously vanishes. No particles are detected as expected.
In case if the detector accelerates uniformly with acceleration $a$   the corresponding Green's function is given by~\cite{Birrell:1982ix}
 \be
D^+ (\Delta\tau')= -\frac{1}{4\pi^2}\sum_k \frac{1}{ \left(\Delta\tau   -i2\epsilon +2i\pi\frac{k}{a}\right)^2}.
\ee
As there are infinite number of poles in the lower -half plane at $\Delta\tau =- 2i\pi\frac{k}{a}$ for positive $k$ the corresponding Fourier transform (\ref{fourier}) leads to the   known   result $\sim  \omega[\exp(2\pi\omega/a)-1]^{-1}$.

In our case the detector- field interaction is described by the combination $(\phi_1-\phi_2)$ rather by a single field $\Phi$ discussed above, see section \ref{interpretation}. Therefore, the relevant response function in our case
is described by the positive frequency   Green's function defined as 
\be
\label{ghost}
 \sim \< 0|\Big(\phi_1(x)-\phi_2(x)\Big),\Big(\phi_1(x')-\phi_2(x') \Big) |0\> , 
\ee
which replaces eq. (\ref{wightman}). One can easily see that this  Green's function given by eq. (\ref{ghost}) identically vanishes as the consequence of the opposite  signs in commutation relations (\ref{comm_2}) and (\ref{comm_1}) describing $\phi_1$ and $\phi_2$ fields, in complete agreement with the arguments  presented in section  \ref{interpretation}. Therefore, the Rindler observer will see an extra energy (\ref{RR}) without detecting any physical particles.
This picture is based, of course, on the standard treatment of gravity as a background field. Such an approximation is justified as long as the produced 
effect  is much smaller than the background field itself. Otherwise, the feedback reaction must be considered. The corresponding analysis, however,  is beyond 
  the scope of this work,  and shall not be discussed here. 
  \section{Topological sectors  and the ghost in 2d QED}
  The main goal of this Appendix is to explain the connection between the description in  terms of the ghost (advocated in the present work) and    the alternative description 
  in terms of subtraction constant (contact term). A short historical detour is warranted here.
  
  The  description in terms of the ghost was advocated by   Veneziano \cite{ven} in the context of the $U(1)_A$ problem,
  while  the alternative description   in terms of subtraction constant (contact term) was developed  by Witten~\cite{witten}. In the Witten's approach  the ghost field does not ever enter the system.  
As long as we work in Minkowski spacetime the two constructions are perfectly equivalent as the subsidiary condition~(\ref{gb}) or~(\ref{gb1}) ensures that the ghost degrees of freedom are decoupled from the physical Hilbert subspace, leaving both schemes with the identical physical spectrum.  In a curved space, on the other hand, we argued that the ``would be'' unphysical ghost  can produce  a positive physical contribution to the energy-momentum tensor (\ref{H2}).  The question arises naturally: where is the corresponding physics hidden in the language of Witten? We refer to section 3.3 of   paper\cite{UZ} where this question has been  elaborated. Here we  just want to mention that the corresponding physics does not go away, but rather, it is  hidden in the boundary conditions.  

This question can be  precisely formulated and answered  in  2d QED in Minkowski space when exact computations, including summation over all topological  sectors can be explicitly performed. 
As we shall see below the summation   over all topological  sectors of the theory   exactly reproduces the contact term (conjectured by Witten) which, on the other hand,  is represented by  the ghost   in the Veneziano approach. We advocate the ghost- based technique   because the corresponding   description   can be easily generalized into curved background, while a similar generalization of the Witten's approach is unknown, and likely to be much more technically complicated, see \cite{UZ} for some comments on this issue. 
    \exclude{We shall discuss   this relation (contact term versus ghost contribution)  by computing the so-called topological susceptibility in flat space background. 
  We use for this purpose a simple 2d QED model when all  essential results and required technique (including summation over all topological  sectors) are known~\cite{SW} because
    all relevant  integrals
  are gaussian and  can be computed  exactly. } We should also note that all formulae in this Appendix are written in Euclidean space where all computations of the path integral (including summation over all topological  sectors) are normally performed. 
   
   Our starting point is the topological susceptibility $\chi$ defined as follows, 
\be
\label{chi}
\chi \equiv \frac{e^2}{4\pi^2} \lim_{k\rightarrow 0} \int \dd^2x e^{ikx}\left< T E(x) E(0) \right> ,
\ee
where $\frac{e}{2\pi}E$ is the topological charge density 
and 
\be
\label{k}
 \frac{e}{2\pi} \int \dd^2x E(x) =k
\ee
is the integer valued topological charge in the 2d $U(1)$ gauge theory, $E(x)=\partial_1A_2-\partial_2A_1$ is the field strength. 
  The  expression for the topological susceptibility in 2d Schwinger model is known exactly~\cite{SW} and it is given by
\be
\label{exact}
\chi= \frac{e^2}{4\pi^2}  \int   \dd^2x \left[ \delta^2(x) - \frac{e^2}{2\pi^2} K_0(\mu |x|) \right] ,
\ee
where $\mu^2=e^2/\pi$ is the mass of the single physical state in this model, and $K_0(\mu |x|) $ is the modified Bessel function of order $0$, which 
is the Green's function of this   massive particle. One can explicitly check that topological susceptibility $\chi$ vanishes in the chiral limit $m \rar 0$
in accordance  with Ward Identities (WI). Indeed, 
\be
\label{chi1}
\chi  &=& \frac{e^2}{4\pi^2}  \int   \dd^2x \left[ \delta^2(x) - \frac{e^2}{2\pi^2} K_0(\mu |x|) \right] \\  \nonumber
&=& \frac{e^2}{4\pi^2} \left[ 1- \frac{e^2}{\pi}\frac{1}{\mu^2}\right]= \frac{e^2}{4\pi^2} \left[ 1-1\right]=0.
\ee
 Important lesson to be learnt  from these calculations    is as follows. Along with the conventional contribution $\sim K_0(\mu |x|)$  from the massive physical state 
 in eq. (\ref{chi1}), there is also a contact term which contributes to the topological susceptibility $\chi$  with the opposite sign.  Without this contribution it would be impossible to satisfy the WI because the physical propagating degrees of freedom can only contribute with sign $(-)$ to the correlation function (\ref{chi1}). As demonstrated  in ref.~\cite{toy}  the contact term    is precisely saturated  by the ghost  $\phi_1$ field\footnote{ One should also remark here that if  the quark's mass  does not vanish $m\neq 0$,  the corresponding WI are automatically  satisfied 
 by the combination of ghost $\phi_1$ field and massive physical field  such that the   right hand side   becomes  proportional to quark's mass $m$ as it should, see \cite{toy} for details.}.
 
 The crucial point relevant for this paper is there existence  of the contact term in (\ref{chi1}) which is present in this correlation function even if one  considers   pure photo-dynamics 
 in 2d without any propagating physical degrees of freedom. This term emerges as a result of the summation over different topological classes  in the 2d pure $U(1)$ gauge theory as we discuss below. The same term   can be computed using the Kogut -Susskind ghost~\cite{KS}  as was shown in \cite{toy}. Both description are equivalent and describe the same physics. One should also recall that the topological susceptibility is related to the $\theta$ dependent portion of the vacuum energy
$ \chi (\theta=0)= - \left. \frac{\partial^2\rho_{\mathrm{vac}}(\theta)}{\partial \theta^2} \right|_{\theta=0}$, and therefore, the sensitivity of $\chi$ to the boundary conditions
automatically implies that the vacuum energy $\rho_{\mathrm{vac}}$ is also very sensitive to the boundary conditions in spite of the fact that the physical Hilbert subspace contains only massive propagating degree  of freedom. 
 
 We follow \cite{SW} and   introduce the  classical ``instanton potential"
in order to describe   the  different  topological sectors of the theory  which are classified  by integer number $k$, see  eq. (\ref{k}). The corresponding configurations   in the Lorentz gauge on two dimensional Euclidean torus   with total area $V$ can be described as follows\cite{SW}:
 \be
 \label{instanton}
 A_{\mu}^{(k)}=-\frac{\pi k}{e V}\epsilon_{\mu\nu}x_{\nu}, ~~~ e E^{(k)}=\frac{2\pi k}{V}, 
 \ee
 such that the action of this classical configuration is
 \be
 \label{action}
 \frac{1}{2}\int d^2x E^2= \frac{2\pi^2 k^2}{e^2 V}.
 \ee
   This configuration corresponds to the   topological charge $k$ as defined by (\ref{k}).
The next step is to  compute   the topological susceptibility for the theory defined by the following partition function
\be
\label{Z}
{\cal{Z}}=\sum_{ k \in \mathbb{Z}}{\int {\cal{D}}}A {e^{-\frac{1}{2}\int d^2x E^2}}.
\ee
   All integrals in this partition function are gaussian and can be easily evaluated using the technique developed in \cite{SW}.
   The result     is determined  essentially  by the classical configurations (\ref{instanton}), (\ref{action}) as real propagating degrees of freedom are not present in the system    of pure $U(1)$ gauge  field theory in two dimensions. We are interested in computing $\chi$ defined by eq. (\ref{chi}). In path integral approach it can be represented as follows, 
   \be\label{chi2}
\chi =\frac{e^2}{4\pi^2} {\cal{Z}}^{-1}\sum_{ k \in \mathbb{Z}}{\int {\cal{D}}}A ~ \int \dd^2x E(x) E(0) ~{e^{-\frac{1}{2}\int d^2x' E^2(x')}}.  
\ee
This gaussian integral can be easily evaluated   using the technique developed in \cite{SW}. The result can be represented as follows, 
   \be\label{chi3}
\chi = \frac{e^2}{4\pi^2} \cdot V\cdot \frac{  \sum_{ k \in \mathbb{Z}} \frac{4\pi^2k^2}{e^2 V^2}  \exp(-\frac{2\pi^2 k^2}{e^2 V})}{ \sum_{ k \in \mathbb{Z}}  \exp(-\frac{2\pi^2 k^2}{e^2 V})}.
\ee
 In the large volume limit $V\rightarrow \infty$ one can evaluate the sums entering (\ref{chi3}) by replacing    $ \sum_{ k \in \mathbb{Z}}\rightarrow \int d k $
 such that the leading term in eq. (\ref{chi3}) takes the form,
     \be\label{chi4}
\chi= \frac{e^2}{4\pi^2} \cdot V\cdot  \frac{4\pi^2}{e^2 V^2} \cdot\frac{e^2 V}{4\pi^2}=  \frac{e^2}{4\pi^2}.
\ee
Few comments are in order. First, the topological sectors with large $k\sim \sqrt{e^2V}$ saturate the series (\ref{chi3}). 
As one can see from the computations presented above, the final result  (\ref{chi4}) is sensitive to the boundaries, infrared regularization, and many other aspects which are normally 
ignored when a  theory from the very beginning is formulated in infinite space with  conventional  assumption about  trivial behaviour at infinity. 
Second, the obtained expression for the topological susceptibility (\ref{chi4}) is finite in the  limit $V\rightarrow \infty$ and  coincides with the contact term from 
exact computations (\ref{exact}) performed for 2d Schwinger model in ref.~\cite{SW}. Third, the result (\ref{chi4})  precisely coincides with Kogut -Susskind ghost 
contribution as demonstrated in ~\cite{toy} and reviewed below (\ref{chi5}).
Therefore, we do observe the sensitivity of $\chi$ (and the vacuum energy $\rho_{\mathrm{vac}}$)  to the far-infrared physics 
  in spite of the fact that the physical Hilbert subspace contains only massive propagating degree  of freedom.

We want to  present one more argument supporting our claim that   in accelerating frame the contact term (which is determined 
in our framework by the ghost contribution) deviates from its  Minkowksi value. The argument is 
based on the Ward Identities when massless fermions are included into the system. As is known, the topological susceptibility (\ref{chi1}) must vanish in the chiral limit $m=0$. As discussed above, it indeed vanishes   as a result of very nontrivial cancellation between the physical contribution with a negative sign and a positive contribution computed  above (subtraction constant)  which is  resulted from the  summation over different topological classes. This subtraction constant in our framework is precisely represented 
by the ghost contribution\cite{toy}.
 Indeed,   the topological density $Q=\frac{e}{2\pi}E$ in 2d QED is given by $\frac{e}{2\pi} E= (\frac{e}{2\pi}){\frac{\sqrt{\pi}}{e}}\left(  \Box\hat\phi -  \Box\phi_1 \right)$\cite{KS}
where $\hat\phi $  is the only physical  massive field of the model with mass $\mu^2=\frac{e^2}{\pi}$ while $\phi_1$ is the Kogut Susskind ghost field.  The relevant correlation function in coordinate space which enters the expression for the topological susceptibility (\ref{chi}) can be explicitly computed as follows, 
\be\label{chi5}
\left< T \!\!\!\right.\!\!\!&&\!\!\!\left.\!\!\! \frac{e}{2\pi} E(x) , \frac{e}{2\pi} E(0) \right> = \\
&=& \left(\frac{e}{2\pi} \right)^2\frac{\pi}{e^2} \int \frac{\dd^2p}{\left(2\pi\right)^2} p^4 e^{-i p x} \left[ - \frac{1}{p^2+\mu^2} + \frac{1}{p^2} \right] 
=    \left(\frac{e}{2\pi} \right)^2 \int \frac{\dd^2p}{\left(2\pi\right)^2}  e^{-i p x} \left[  \frac{p^2}{p^2+\mu^2}  \right] \nonumber\\
&=&   \left(\frac{e}{2\pi} \right)^2 \int \frac{\dd^2p}{\left(2\pi\right)^2} e^{-i p x} \left[ 1 - \frac{\mu^2}{p^2+\mu^2} \right]  
=   \left(\frac{e}{2\pi} \right)^2\left[ \delta^2(x) - \frac{e^2}{2\pi^2} K_0(\mu |x|)\right] \, , \nonumber
\ee
where we used the known expressions for the  Green's functions (the physical massive field $\hat\phi $ as well as the ghost  $\phi_1 $ field).  The obtained expression is precisely the  result (\ref{chi1}), as anticipated.  Our additional argument supporting the main  claim (that   in accelerating frame the contact term    represented by $\delta^2(x)$ function in (\ref{chi5}) is different from its  Minkowksi value)  goes as follows. The contribution of the physical massive state represented by $K_0(\mu |x|)$ in (\ref{chi5}) obviously changes
when we go from Minkowski space to  the accelerating frame  such that the corresponding massive Green's function as well as the residue  $\left(\frac{e}{2\pi} \right)^2$ would generally depend  on acceleration $``a"$. It obviously implies that the ghost contribution represented by $\delta^2(x)$ function in (\ref{chi5})
must also depend on on acceleration $``a"$. This is because   the Ward Identity $\int\dd^2x \left< T \frac{e}{2\pi} E(x) , \frac{e}{2\pi} E(0) \right>  =0$ must be respected in the accelerating frame. 
The WI can  only be satisfied in accelerating frame if the corresponding subtraction contribution (\ref{chi4}) does  depend  on acceleration 
$``a"$. But in our framework this contribution is precisely determined by the KS ghost (\ref{chi5}).
 This argument supports our claim that  the KS ghost contribution (and  the vacuum energy, correspondingly) should generically depend on acceleration $``a"$\footnote{ In order to explicitly test this argument, one should  repeat the path integral calculation of refs \cite{SW} in accelerating frame, including the infrared regularization. This subject is beyond the scope of the present analysis. As we mentioned above, we do not know presently  how to proceed with the computations similar to (\ref{chi4})   in the accelerating frame.}.

$\bullet$ The most important lesson to be learnt from  these simple computations in this simple  model  is that the dynamics of gauge  
systems is quite sensitive to the boundary conditions.
  Therefore, when such a system is promoted to a curved or time dependent background, it is quite naturally to expect that the vacuum energy will be sensitive to the properties of this  background.   We advocate the ghost- based technique  to account for this physics because the corresponding   description   can be easily generalized into curved background, while a similar  generalization  (without the ghost, but with explicit 
accounting for the infrared behaviour at the boundaries) is unknown and  likely to be much more technically complicated. In different words, the description in terms of the ghost is a matter of convenience to  (effectively) account for  the far- infrared  effects  in topologically nontrivial sectors of the theory.

\end{document}